\documentclass[acmtog]{acmart}
\renewcommand\footnotetextcopyrightpermission[1]{}
\setcopyright{none}
\AtBeginDocument{%
  \providecommand\BibTeX{{%
    \normalfont B\kern-0.5em{\scshape i\kern-0.25em b}\kern-0.8em\TeX}}}

\newcommand{\etal}{\emph{et al.}}

\usepackage[normalem]{ulem}
\newcommand{\myremove}[1]{}
\newcommand{\mysecondremove}[1]{}

\newenvironment{new}{\color{black}}\ignorespacesafterend 
\newenvironment{sjnew}{\color{black}}\ignorespacesafterend 
\ignorespacesafterend 
\ignorespacesafterend 

\newenvironment{move}{\color{black}}

\usepackage{booktabs}
\usepackage{pifont}
\newcommand{\cmark}{\ding{51}}%

\usepackage{url}
\usepackage{textcomp}
\usepackage{hyperref}

\begin{document}

\title{Foveated Rendering: Motivation, Taxonomy, and Research Directions}

\author{Susmija Jabbireddy}
\affiliation{%
  \institution{University of Maryland, College Park}
  \city{College Park}
  \country{USA}}
\email{jsreddy@umd.com}
\author{Xuetong Sun}
\affiliation{%
  \institution{University of Maryland, College Park}
  \city{College Park}
  \country{USA}}
\author{Xiaoxu Meng}
\affiliation{%
  \institution{University of Maryland, College Park}
  \city{College Park}
  \country{USA}}
  
\author{Amitabh Varshney}
\affiliation{%
  \institution{University of Maryland, College Park}
  \city{College Park}
  \country{USA}}

\begin{abstract}
With the recent interest in virtual reality and augmented reality, there is a newfound demand for displays that can provide high resolution with a wide field of view (FOV). However, such displays incur significantly higher costs for rendering the larger number of pixels. This poses the challenge of rendering realistic real-time images that have a wide FOV and high resolution using limited computing resources. The human visual system does not need every pixel to be rendered at a uniformly high quality. Foveated rendering methods  provide {\em perceptually} high-quality images while reducing computational workload and are becoming a crucial component for large-scale rendering. In this paper, we present key motivations, research directions, and challenges for leveraging the limitations of the human visual system as they relate to foveated rendering. 
We provide a taxonomy to compare and contrast various foveated techniques based on key factors. 
We also review aliasing artifacts arising due to foveation methods and discuss several approaches that attempt to mitigate such effects. 
Finally, we present several open problems and possible future research directions that can further reduce computational costs while generating perceptually high-quality renderings.

\end{abstract}

\ccsdesc[500]{General and reference~Surveys and overviews}
\ccsdesc[300]{Computing methodologies~Mixed / augmented reality}
\ccsdesc[500]{Computing methodologies~Rendering}
\ccsdesc[100]{Computing methodologies~Antialiasing}
\ccsdesc[300]{Computing methodologies~Perception}
\ccsdesc[100]{Computing methodologies~Ray tracing}

\keywords{foveated rendering, foveated displays, eye-tracking, gaze-contingent rendering, variable-rate sampling, head mounted displays}

\maketitle

\section{Introduction}
\label{Introduction}

Virtual and augmented reality (VR and AR) are poised to transform computer-mediated communication spanning an exciting range of applications in science, engineering, medicine, arts, entertainment, commerce, and several other areas. While impressive, the current-generation VR and AR systems are unable to match the visual fidelity of our real-world experiences along several dimensions, including resolution, the field of view, and latency. 
Delivering such a high-quality visual experience in real-time requires enormous computational resources beyond the recent improvements in the hardware and software systems for rendering. 

A very small fraction of the scene that is projected on the fovea, the center of the human retina, is perceived by the human visual system (HVS) at its finest details. 
Visual acuity is the ability to observe detail and is measured as the grating resolution in cycles/degrees. Studies show that though HVS has a wide field of view (FOV), the region that has the highest visual acuity, also known as the foveal region, covers only the central $2.5$\textdegree\ of the visual field~\cite{levin2011adler}. Foveation refers to a decrease in the acuity with angular distance in the human visual system \cite{guenter2012foveated}. Foveated rendering leverages this feature of the human visual system to selectively render a small fraction of the graphics frame in fine detail. 
Foveated rendering allows us to reconcile the mutually conflicting goals of high visual realism, interactive frame rates, and low power consumption on modern VR and AR devices.
As foveated rendering becomes an integral part of future AR/VR devices, it is important to understand the current state and the observed trends in the field.

\begin{figure}
\centering
\includegraphics[width=0.45\textwidth]{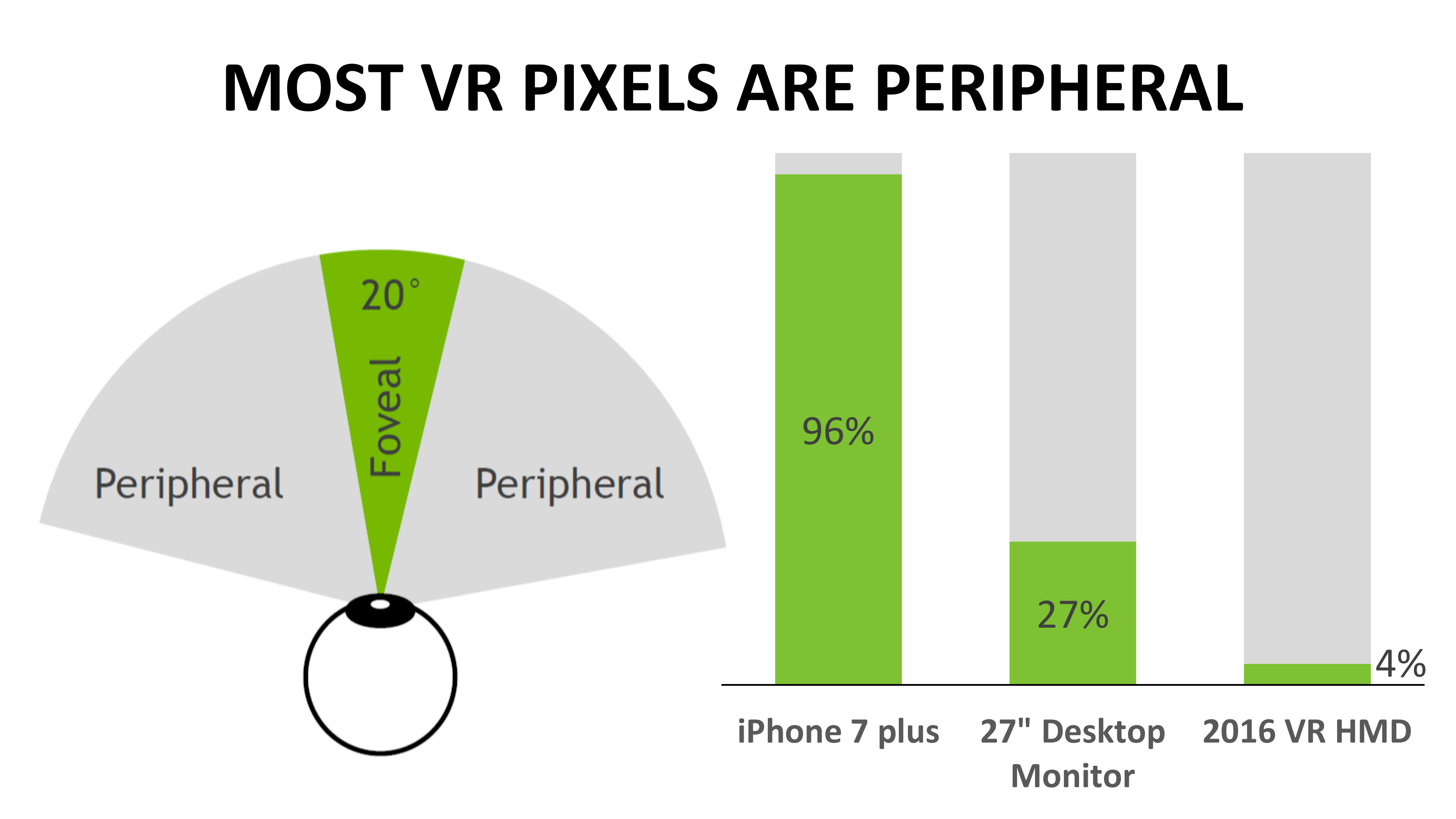}
\caption{Percentage of pixels that lie in the fovea region when viewed on an iPhone 7 plus screen, 27" desktop monitor and a VR head-mounted display. Image adapted from Patney~\etal~\shortcite{patney2016towards}.}
\label{pixels}
\end{figure}

The concept of foveated displays is not new. Early implementations of the gaze-contingent displays find applications in building flight simulators~\cite{murphy_lod}. However, early methods were developed mainly for desktop monitors, where the FOV was approximately $43$\textdegree\ $\times$ $33$\textdegree~\cite{murphy_lod}. This is a fraction of modern wide FOV displays. 
Patney~\etal~\shortcite{patney2016towards} give a comparison of the percentage of pixels that lie in the peripheral region across various devices ranging from a small FOV smartphone screen to a wide FOV VR HMD, as shown in Figure~\ref{pixels}. Thus, foveated rendering becomes even more important for VR and AR headsets compared with traditional displays. In a $100$\textdegree\ wide FOV HMD, only 4\% of the screen pixels lie in the foveal region while the rest lie in the peripheral region~\cite{patney2016towards}. When such a large percentage of pixels lie in the low acuity peripheral region, efficient rendering systems can obtain significant computational gains by allocating fewer resources to render the peripheral region.

Until recently, real-time eye-tracking was not commonly available in VR and AR headsets. However, with the commoditization of eye-trackers in VR and AR headsets, foveated rendering techniques can use the gaze information to dynamically render the foveal region at a higher quality and the rest at a lower quality without a drop in the overall perceived quality of the image. As the FOV of future HMDs increases to match that of the human vision, foveated rendering will emerge as an essential component of the real-time rendering 3D graphics pipeline for large-scale, wide FOV and high-resolution VR and AR displays.
4D light fields is another area that can benefit from the foveated rendering techniques. 4D light fields represent a scene from multiple camera positions, which incur a huge rendering cost. Sun~\etal~\shortcite{sun_lf} and Meng~\etal~\shortcite{meng_lf} present methods that use foveated rendering techniques to accelerate the rendering process of light fields, thus enabling real-time light field rendering.

Recently, a classification of foveated displays has been presented by
Spjut~\etal~\shortcite{foveated_display_classification}. 
Their classification categorizes foveated rendering systems based on the resolution distribution function and gaze contingency.
The resolution distribution function measures how well the system's acuity distribution matches the pre-measured visual acuity for the user (as measured by the Snellen eye chart). Gaze contingency measures how the system adapts to changes in the gaze direction. 
This paper focuses on foveated rendering methods for VR/AR systems. We consider different attributes that characterize foveated rendering, including several visual factors, resolution distribution, foveation space, gaze information, and anti-aliasing response. 
Further, we have classified several influential research papers along these dimensions.
A similar review is also provided in a concurrent work by Mohanto~\etal~\cite{MOHANTO2022474} with a different taxonomy of foveated rendering methods. 

We organize our paper as follows. Section~\ref{Human Visual System} provides a brief overview of the human visual system and its limitations that enable foveated rendering. A taxonomy of various foveated rendering methods based on their salient characteristics is presented in Section~\ref{Taxonomy}. Section~\ref{Evaluation} discusses the methods used to evaluate the perceived quality of the foveated image. Finally, some concluding remarks along with the challenges and possible future research directions are presented in Section~\ref{Summary}.

\section{A brief overview of Human Visual System}
\label{Human Visual System}

\begin{figure}[!h]
\centering
\includegraphics[width=0.4\textwidth]{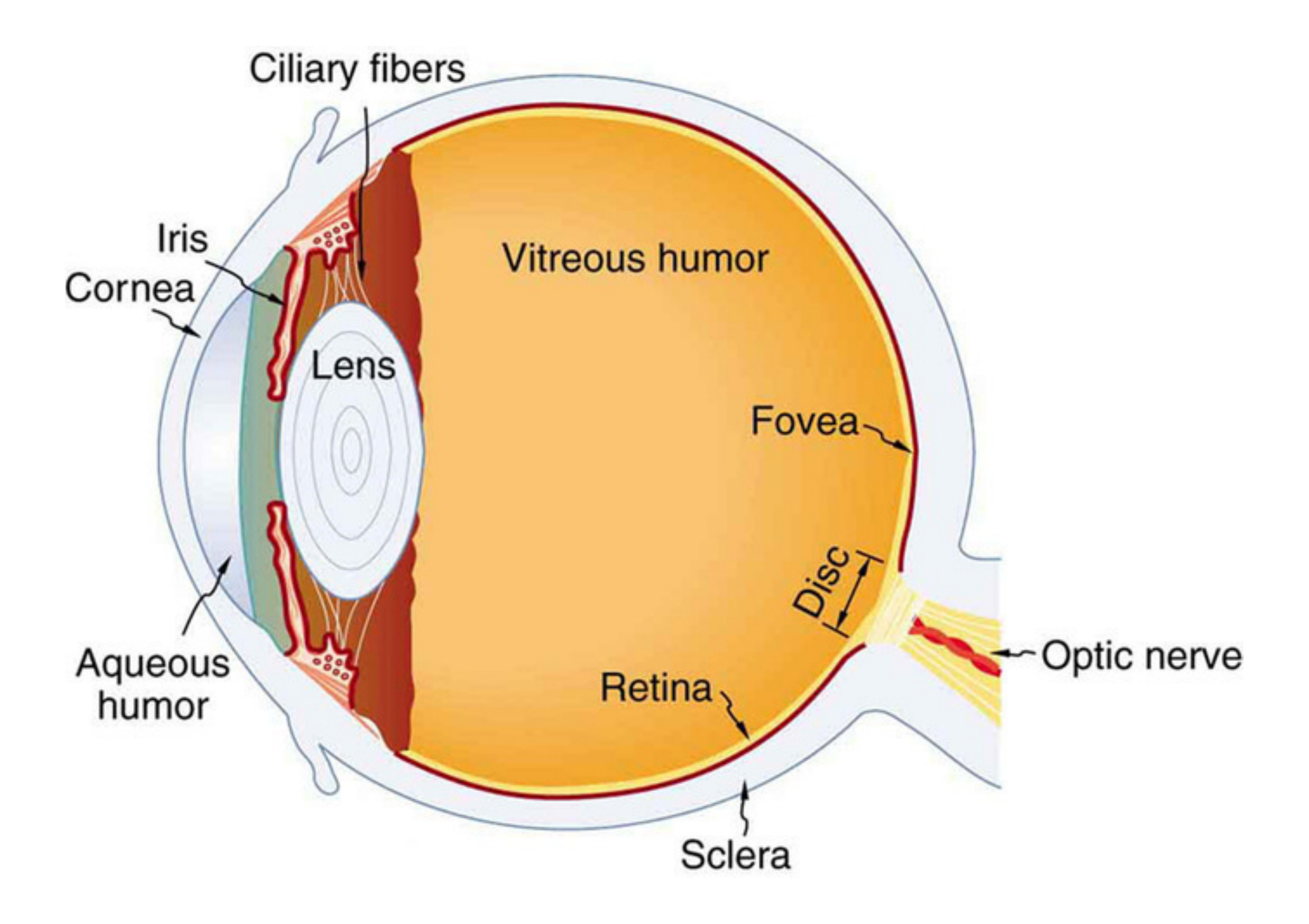}
\caption{Anatomy of the human eye. Image from Urone~\etal~\shortcite{physics_of_eye}, reproduced under Creative Commons Attribution License.}
\label{Human eye anatomy}
\end{figure}

Understanding human perception and its limitations is beneficial to improve the rendering quality and efficiency. 
The eye and the brain make up the human visual system. While the eye serves as a camera, the brain is responsible for processing the information.
When a light ray hits the eye, it first passes through the cornea and undergoes refraction.
The refracted ray passes through the aqueous humor, the iris, the lens, the vitreous humor, and finally, reaches the retina - the image sensor of the human eye. 
Figure \ref{Human eye anatomy} shows the complete anatomy of the human eye.
The light photons reaching the retina are then detected and converted to electrical signals by photoreceptors.
The information from the photoreceptors is transmitted to the central neural system by ganglion cells for further processing.

\begin{figure}[!h]
\centering
\includegraphics[trim= 0 0 0.2cm 0cm, clip, width=3in]{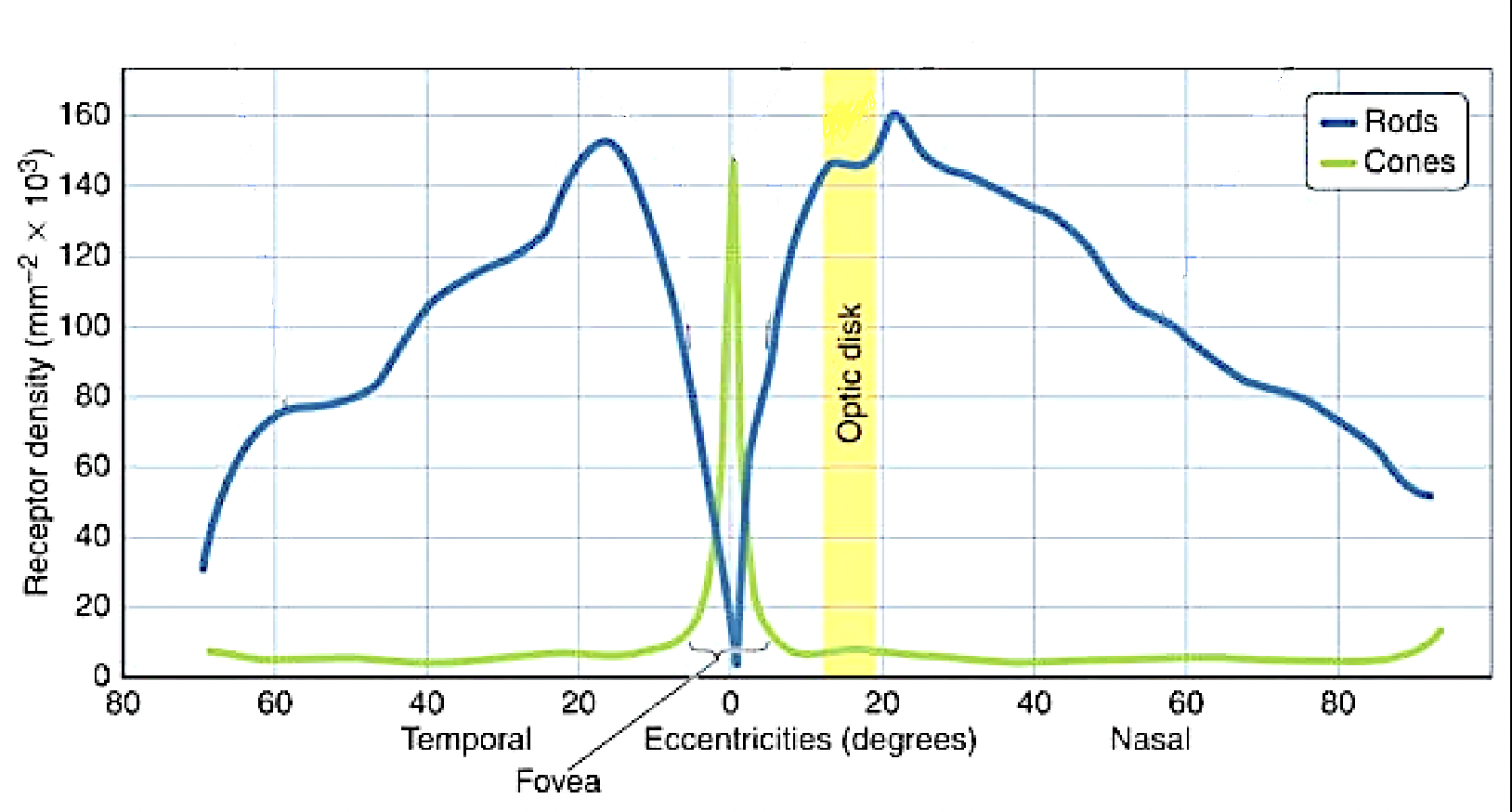}
\caption{Density distribution of photoreceptors (rods and cones) with eccentricity in the human eye. Image adapted from Mustafi~\etal~\shortcite{rods_cones_distribution}}
\label{rods and cones}
\end{figure}

It is well known that not all the information that enters the eye reaches the brain. 
A significant amount of compression occurs at the retinal level.
Several components are responsible for this pre-processing.
Firstly, the distribution of photoreceptors and the ganglion cells is not uniform across the retina. 
There are two categories of photoreceptors, namely rods and cones.
The rods and cones differ in characteristics such as shape, distribution across the retina, patterns of synaptic connections, among many others~\cite{neuroscience_book}.
We have 100's of millions of rods, while the number of cones is around six-seven million~\cite{rodscones_taylor}.
Moreover, a large portion of the cones is concentrated near the central part of the retina, known as the fovea, and their density greatly decreases as the distance from the center increases. 
On the other hand, rods are spread out across the entire retina and are absent at the fovea. Figure~\ref{rods and cones} shows the density distribution of rods and cones across the visual field. 
In addition, rods are highly sensitive to light, whereas cones provide excellent spatial discrimination and are responsible for color vision.
The differences in the rod and cone systems characterize three modes of vision based on the amount of light that reaches the human eye. 
The first one is scotopic or night vision, which facilitates perception in dim lighting. In this type of vision, only rods are active. Rods cannot discriminate colors, and therefore, scotopic vision is a grayscale vision. The second type of vision is photopic vision, or daylight vision, which provides for color perception mediated by the cone cells. Mesopic vision, the third type of vision, is a combination of photopic and scotopic vision and is perceived in low but not quite dark lighting conditions. Both rods and cones are simultaneously active for this vision. 
Most indoor and outdoor lighting situations use photopic vision, where cones are the most active ones. However, if one wishes to render dim lighting, the properties of rods are to be considered. 
Therefore, the perceived level of detail at any point in time also depends on the active mode of vision.

The ganglion cells in the inner part of the retina transmit the information from the photoreceptors to the neural processing system. While we have 100's of millions of photoreceptors, the number of ganglion cells present in the retina is only around 1.2 million~\cite{retinal_ganglion}. 
This many-to-one connection further compresses the information reaching the brain.
Studies~\cite{ganglion_cell_density} observed that the ganglion cell density is also non-uniform across the retina. The number of photoreceptors connected to a single ganglion cell increases with an increase in the distance from the fovea.

The spatially varying photoreceptor and the ganglion cell densities, with a higher density at the center, lead to foveated vision in the human visual system. As a result, the visual sensitivity is maximum at the fovea and reduces towards the periphery. 
The exact transition point from the foveal to the peripheral region is not defined uniformly across various disciplines.
The eccentricity of a point in the retina is defined as the angular distance of that point from the fovea. The fovea is circumscribed by para-fovea, which is circumscribed by peri-fovea.
Based on these regions, the human visual field can be broadly divided into central and peripheral vision. Neuro-psychologists typically define central vision as the region from $0$\textdegree\ to $2.6$\textdegree\ - $3.6$\textdegree\ eccentricity, with everything beyond that considered as peripheral vision~\cite{loschky2019contributions}. Visual cognition researchers generally consider the region from 0\textdegree\ to 5\textdegree\ eccentricity as the central vision, comprising the foveal region from $0$\textdegree\ to approximately $0.5$\textdegree\ - $1$\textdegree\ eccentricity and para-fovea from approximately $0.5$\textdegree\ - $1$\textdegree\ to $5$\textdegree\ and anything beyond 5\textdegree\ is considered peripheral vision.  Vision science researchers consider the central vision to extend from 0\textdegree\ to 10\textdegree\ with the peripheral vision beyond that~\cite{loschky2019contributions}. Whatever be the exact definition, the vast majority of the human visual field falls in peripheral vision.

Foveated rendering exploits the foveation in the human visual system and renders the peripheral region at a relatively lower quality. 
For a more comprehensive survey on the physiological and perceptual aspects of human vision and limitations of the human visual perception, readers are referred to the work by Weier~\etal~\shortcite{weier2017perception_driven}.

\section{A Taxonomy of Foveated Rendering Systems}
\label{Taxonomy}

\begin{figure*}[h]
\centering
\includegraphics[width=0.9\linewidth]{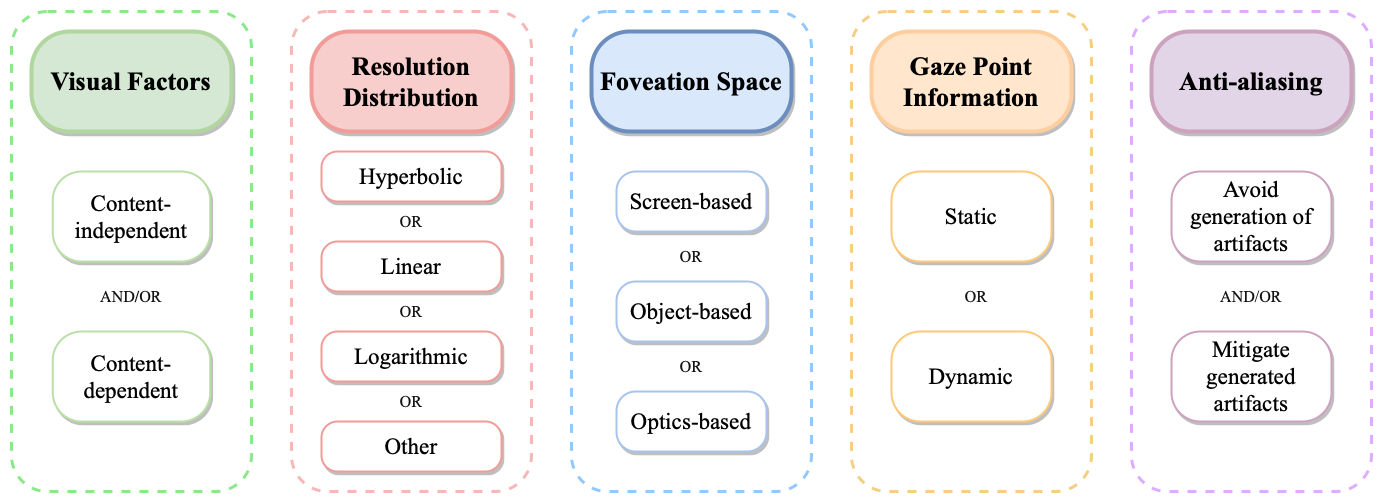}
\caption{Taxonomy of the foveated rendering methods based on some of the key factors. A full foveated rendering pipeline might choose any sequence of options from these categories.} 
\label{fig_taxonomy}
\end{figure*}

The non-uniform distribution of rods and cones across the retina, along with their varying responses to the incoming light, leads to several interesting properties that change from the foveal to the peripheral vision.
For instance, perception of detail reduces with an increase in eccentricity and the time to perceive a same-sized object increases with eccentricity~\cite{sun_perception}. 
Rods have limited spatial discrimination while cones have excellent spatial localization and color discrimination~\cite{neuroscience_book}.
However, it is interesting that the ability to detect motion in the peripheral region is higher than what would be expected. Moreover, the ability to differentiate an object's relative velocity and flicker sensitivity stays uniform across the visual field.
In addition, Patney~\etal~\shortcite{patney2016towards} show that preserving contrast in the peripheral region improves image perception for filtered images.
Studies also show that while color perception decreases with an increase in eccentricity, the decrease is gradual and color discrimination survives at high eccentricities~\cite{color_degradation}.
Rosenholtz~\cite{ruth_peripheral} argues that peripheral vision is not a mere low-resolution version of foveal vision, but instead is responsible for a texture like representation that preserves the summary statistics of a scene. 
The different properties that are needed to be preserved in the peripheral region depends on the task at hand.
Foveated rendering can leverage these human vision characteristics that vary with eccentricity.

While most real-time rendering uses the rasterization pipeline, ray-tracing provides more realistic images with accurate reflections, refractions, and shadows and is gradually eliciting greater interest~\cite{ludvigsen2010real}. Foveation can be applied to both rasterization and ray-tracing pipelines.
The graphics pipeline for real-time rasterization is often dominated by the computational complexity of shading operations. 
Shading costs can be reduced by simplifying the operation, pre-computing parts of the operation, sharing results across a set of fragments, and/or reducing the number of operations.
Foveated rendering attempts to reduce the average number of operations per pixel by leveraging the non-uniform visual sensitivity across the visual field. 
We next categorize and compare foveated rendering techniques based on their characteristics and discuss their use in efficient real-time rendering. 
Figure ~\ref{fig_taxonomy} gives an overview of the presented taxonomy.

Section~\ref{Visual Attributes} presents various visual attributes of the human visual system that spatially vary across the visual field.
Section~\ref{Human Visual Models} discusses various models representing the spatial variance of visual sensitivity,
and Section~\ref{Foveation space} considers the stage at which foveation is implemented in foveated rendering systems.
Section~\ref{Static vs Dynamic} gives an overview of the techniques based on their relation to gaze tracking. 
Finally, Section~\ref{Anti-Aliasing} discusses anti-aliasing techniques for foveated rendering.

\subsection{Visual Factors}
\label{Visual Attributes}
Visual perception depends on various factors including, but not limited to, eccentricity, contrast, luminance, brightness, color, shape, motion, fixations, and saliency. 
In this section, we survey the foveated rendering methods based on the different visual factors considered.


\subsubsection{Content-independent factors} 
\label{Spatial Resolution}
One widely used visual measure is visual acuity, defined as the reciprocal of the angle (in minutes) subtended by a just resolvable region at a given eccentricity. 
Visual acuity varies across the visual field. 
Early works~\cite{ANSTIS1974589} show that acuity drops with an increase in eccentricity. 
Therefore, reducing the resolution as a function of angular distance from the gaze point has always characterized foveated rendering.
One of the distinguishing features among the various foveated rendering approaches has been the way this reduction is accomplished. 
One common way is to render multiple discrete layers independently; each sampled at a different rate~\shortcite{guenter2012foveated, swafford2016user}. These layers, called eccentricity layers, are later upsampled to the screen resolution and composed together.
As shading is the most time-consuming operation in the rendering pipeline, certain works~\shortcite{he2014extending, vaidyanathan2014coarse, nvidia_multiresolution_shading} vary the shading rate across the visual field, shading only once per several pixels in the peripheral region.
Some recent works~\cite{meng2018kernel, visual_polar_space, tursun2019luminance} also demonstrate varying resolution in a more continuous manner. 
These methods map the visual field to a non-linear space where a uniform rendering matches the visual acuity of human vision.
While the above methods are mainly developed using a rasterization pipeline, ray tracing methods naturally allow for smooth non-uniform sampling in the screen space. Weier~\etal~\shortcite{raytracing_hmd} and Fujita and Harada~\shortcite{raytracing_vrheadset} incorporate foveated rendering into ray-tracing and achieve sparse sampling by varying the sampling probability with an increase in distance from the gaze point.
For more discussion on the trade-offs between discrete and continuous distributions, we refer the readers to the recent work on foveation displays~\cite{classification_foveated_displays}.
The idea behind these methods is to mimic a distribution that closely aligns with the visual acuity of the HVS. Different methods approximate the non-uniform sampling distribution function differently to best match acuity across the visual field; we discuss this in detail in Section~\ref{Human Visual Models}.

Furthermore, recent studies~\cite{barbot_asymmetries, jared_asymmetries} show that the visual acuity is higher across the horizontal axis than the vertical axis at the same eccentricity values. Studies further show that the lower visual field has better visual acuity. 
Most of the current foveated rendering techniques do not factor this asymmetry during rendering and assume uniform acuity across horizontal and vertical directions.
One exception is the recent method by Ye~\etal~\shortcite{rmfr} that processes the horizontal and vertical visual fields independently, achieving a superior image quality.


\subsubsection{Content dependent factors}
\label{Contrast Sensitivity}

\paragraph{Contrast}
Image contrast is another major factor impacting visual perception. 
The minimum amount of contrast (contrast threshold) required to detect a pattern depends on its spatial frequency, in addition to the content-independent eccentricity.  
The reciprocal of contrast threshold is called contrast sensitivity, and it is characterized by the contrast sensitivity function (CSF) across the human visual field.
The function describes the capacity of the HVS to recognize differences in patterns as a function of spatial frequencies~\cite{CSF_Book}. 
Similar to spatial acuity, contrast sensitivity for a given spatial frequency is highest at the fovea and decreases with increasing eccentricity. 
The contrast sensitivity depends on several factors such as size, contrast, and the viewing angle of the target pattern, and several functions for modeling CSF have been proposed in the literature~\cite{csf_lowfreq, csf_formula}.

\begin{figure}[!h]
\centering
\includegraphics[width=0.45\textwidth]{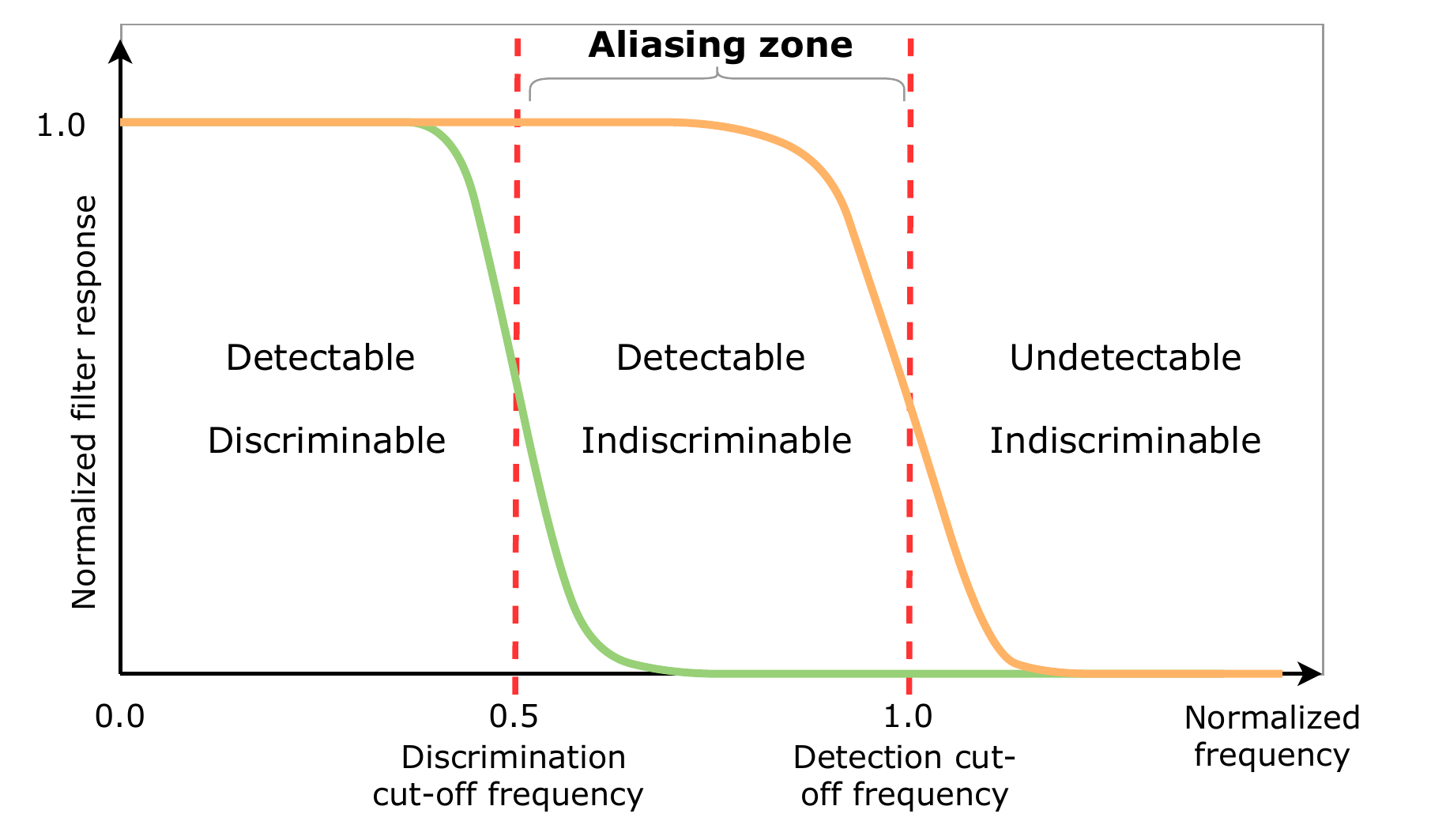}
\caption{An illustration of the presence of aliasing zone in human peripheral vision. Bandlimiting filter response is plotted on the y-axis against the angular frequency on the x-axis, normalized to the detection cut-off frequency. Aliasing zone represents the range of frequencies that lie between the discrimination and detection cut-off frequencies.
Image adapted from Patney~\etal~\shortcite{patney2016towards}.}
\label{fig: Detection Resolution}
\end{figure}

The relationship between contrast sensitivity and spatial frequency varies across the visual field~\cite{csf_foveal_peripheral}.
In general, the foveal CSF shows a gradual loss in contrast sensitivity with increasing spatial frequency, while peripheral CSF shows that contrast sensitivity drops abruptly. The contrast sensitivity function has a cut-off spatial frequency beyond which a higher frequency is not discernible. The cut-off spatial frequency depends on the eccentricity of the image patch location in the visual field. 
Further, the contrast sensitivity function is task-dependent.
The peripheral CSFs for detection and discrimination tasks indicate different limits to performance at high spatial frequency~\cite{csf_resolution_detection}.
The peripheral contrast sensitivity for the discrimination task is observed to degrade at a much faster rate than for the detection task, as demonstrated in Figure~\ref{fig: Detection Resolution}.
The difference between the cut-off frequencies for the detection task and the discrimination task indicates that the human visual system can perceive higher frequencies in the periphery but cannot resolve details. 
The range of frequencies between the detection and discrimination cut-off frequencies is known as the aliasing zone.
The foveated renderers that determine the sampling rate based on the discrimination cut-off frequency provide faster performance than those based on the detection cut-off frequency. 
However, rendering the peripheral region based on the discrimination cut-off frequency reduces contrast for regions whose frequencies are in the aliasing zone. 
To maintain the detection ability, Patney~\etal~\shortcite{patney2016towards, patney_perceptually} suggest enhancing peripheral contrast in the aliasing zone, which was degraded by filtering to maintain the perceptual quality of a non-foveated image.
Tursun~\etal~\shortcite{tursun2019luminance} exploits the influence of luminance contrast on visual perception.
They observe that while a given reduction of spatial resolution in high-contrast regions reduces the perceived quality, the same reduction in low-contrast regions maintains it.
The minimal required resolution for each region is estimated using the local luminance contrast and the angular distance from the gaze point.
In addition to luminance, the sensitivity to color also reduces towards the peripheral region. The hue resolution or the number of gray levels within each RGB channel that can be perceived reduces with eccentricity~\cite{color_discrimination}.
Liu~\etal~\shortcite{sheng_color} show that the number of bits representing each color channel can be monotonically reduced from 8 to 4 as the eccentricity increases from 0\textdegree~to 30\textdegree. 

\paragraph{Saliency}
Visual saliency, often a good indicator of visual attention, is another factor influencing the perceived level of detail.
Visual saliency is a distinct subjective quality where certain regions of a scene are more likely to draw the viewer's attention when compared to their surrounding regions.
The concept of foveal and attentional spotlights states that the attention and gaze do not necessarily coincide~\cite{levin2011adler}. 
As the user's attention is automatically attracted to the visually salient stimuli, identifying such salient regions can redistribute the rendering cost by allocating more resources to the visually important regions.
Many computational models have been introduced to estimate visual saliency~\cite{mesh_saliency_eye_fixations, mesh_saliency, mesh_saliency_spectral_processing, saliency_lowres_gray}. 
In addition to the low-level features like local color and brightness contrast, high-level features such as the position and identity of the objects and the scene context have a significant influence on the visual attention model.
The set of visually salient regions is not uniquely considered across various research works.  
For example, the adaptive multi-rate shading method by He~\etal~\shortcite{he2014extending} considers regions near object silhouettes, shadow edges, and regions of potential specular highlights as visually salient. 
A higher sampling rate is used to render these regions so that they are perceived at relatively higher acuity.
On the other hand, Stengel~\etal~\shortcite{stengel2016adaptive} consider the regions of high spatial and temporal contrast and saturated colors as visually significant to develop a foveated rendering system.



\subsection{Eccentricity-based analytical visual models}
\label{Human Visual Models}
Foveated rendering systems vary resolution or sampling rate across the visual field based on one or more visual factors discussed above. 
\begin{sjnew}
Based on the scene content and task, these systems approximate the human visual field using analytical expressions.
\end{sjnew}
\mysecondremove{
These methods use analytical approximations to model visual sensitivity fall-off from the foveal to the peripheral region.}
These approximations allow different kinds of pixel distributions. In this section,
\mysecondremove{
we categorize the foveated rendering systems based on the nature of different fall-off distributions.}
\begin{sjnew}
we discuss the similarities and differences among the different distributions.
\end{sjnew}

\subsubsection{Hyperbolic model}
\label{CMF Linear Models}
Visual acuity can be quantitatively represented as the reciprocal of the minimum angle of resolution~\cite{weymouth}, which is the smallest angle at which two points are perceived as different. 
The minimum angle of resolution increases linearly with an increase in eccentricity.
\begin{sjnew}
One possible mathematical characterization of visual acuity over eccentricity is given by
\end{sjnew}

\mysecondremove{
The mathematical characterization of minimum angle of resolution $\omega$ with eccentricity $e$ is given by the equation:}
\begin{equation}
\text{Acuity} = \frac{1}{m e + \omega_{0}}
\label{CMF_eq}
\end{equation}
where, $\omega_{0}$ corresponds to the smallest resolvable angular resolution that occurs at the fovea ($e=0$) and $m$ represents the slope. 
\begin{sjnew}
We refer to this distribution as the hyperbolic model where the acuity changes as a function of 1/eccentricity.
\end{sjnew}
This hyperbolic model serves as a good approximation for visual acuity at low eccentricities (less than $8$\textdegree\ angular radii)~\cite{guenter2012foveated}, after which the acuity drops more steeply. 
Many foveated renderers ~\cite{guenter2012foveated,swafford2016user, patney2016towards, vaidyanathan2014coarse, he2014extending, stengel2016adaptive} base their sampling distribution on this hyperbolic fall-off of acuity.

\begin{figure}[!h]
\centering
\includegraphics[trim=1cm 1cm 1cm 1cm, clip, height=4in]{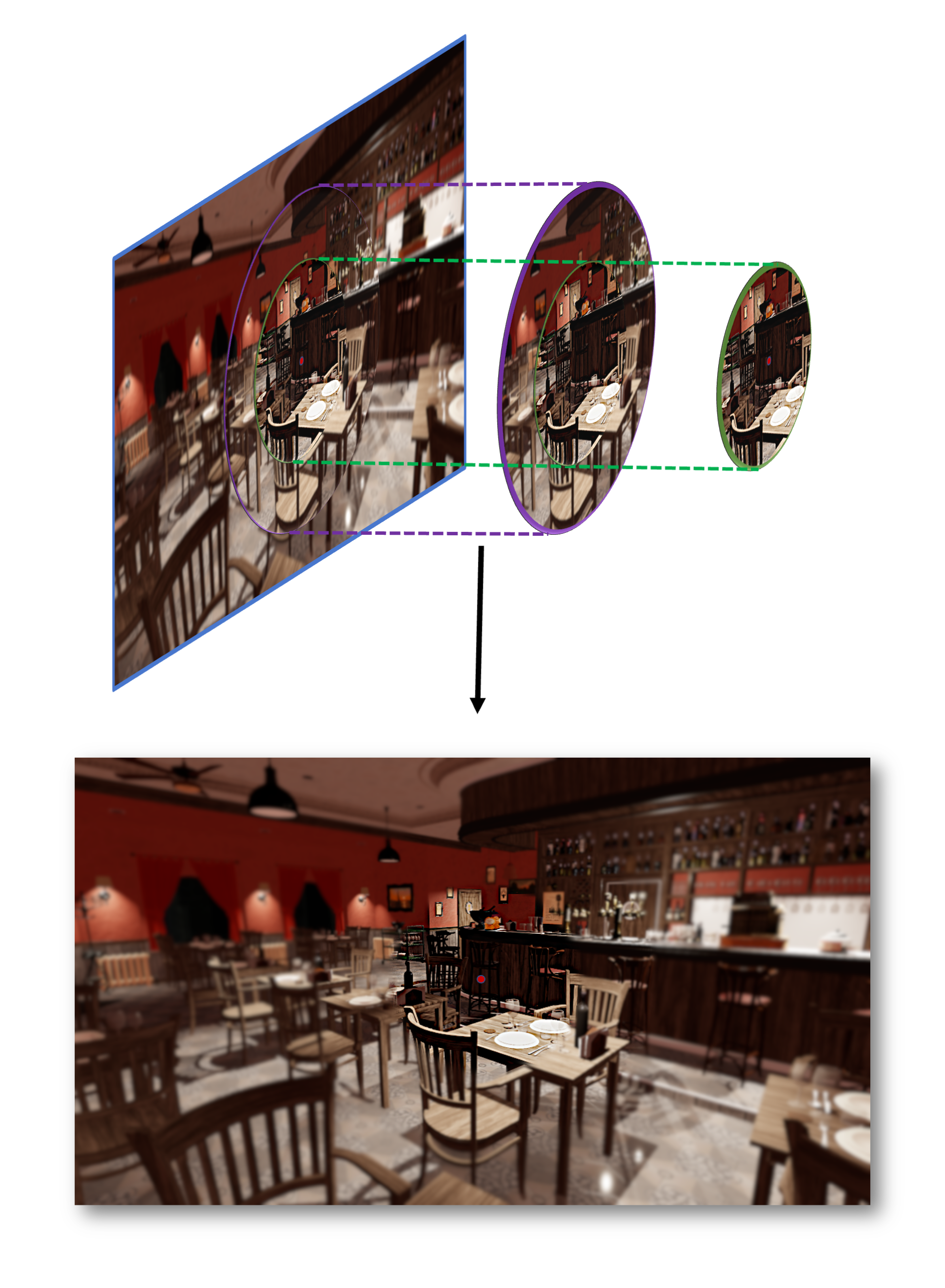}
\caption{Three discrete layers are used to match the human acuity across the visual field. The three layers are rendered at different resolutions based on the distance to the gaze point (red dot). Image adapted from Guenter~\etal~\shortcite{guenter2012foveated}.}
\label{Foveated 3D Graphics}
\end{figure}

Guenter~\etal's~\shortcite{guenter2012foveated} pioneering work shows that significant performance improvement is possible with foveated rendering for rasterization. Assuming that a few discrete layers are sufficient to model the acuity fall-off in the human visual system, they use three nested and overlapping rectangular layers rendered at different resolutions, as shown in Figure~\ref{Foveated 3D Graphics}.
These three layers, known as eccentricity layers, are centered at the gaze point or the fovea. The innermost, foveal layer, is rendered at the highest resolution (which is the native display resolution). The middle layer is larger than the inner layer and is rendered at a relatively lower resolution. The outermost layer covers the entire screen and is rendered at a much lower resolution. All three layers are then interpolated to the display resolution and blended smoothly. The hyperbolic function in Equation~\ref{CMF_eq} is used to compute the size and resolution for the eccentricity layers. The parameters depend on the minimal angular resolution slope $m$, which is estimated from the user studies.
The approach assumes symmetric radial acuity fall-off, ignoring the differences in the horizontal and vertical axes of the visual field. They limit their method to three layers to approximate the human visual system, although more layers can give a better approximation at the cost of increased complexity. 
Thus, the the total number of rendered pixels is reduced compared to rendering the entire frame at a uniformly high resolution. 

Stengel~\etal~\shortcite{stengel2016adaptive} extend the above model to incorporate the effect of smooth-pursuit eye movements. This type of eye movement is triggered unconsciously when a moving object attracts attention. They model the region of focus as a straight line, obtained as integration of gaze positions over consecutive frames. 
The eccentricity is now measured based on the distance to this line. Consequently, the rendering at the highest-resolution is over a larger elliptical region enveloping the straight line, instead of a circular region around a single point. 
This method gives superior performance on higher latency displays.

Spjut~\etal~\shortcite{foveated_display_classification} characterize the acuity distribution of HVS using a hyperbolic model and evaluate a foveated display based on how well it matches this distribution. For natural viewing of images, the hyperbolic model has proven to be accurate for small eccentricity values~\cite{stengel2016adaptive, guenter2012foveated}. However, often, the region beyond $30$\textdegree\ eccentricity is rendered at a uniform lower resolution, which potentially limits the extent of speedup that can be achieved through foveated rendering.

\subsubsection{Linear model}
\label{Linear fall-ff}
\begin{figure}[!h]
\centering
\includegraphics[width=3.5in]{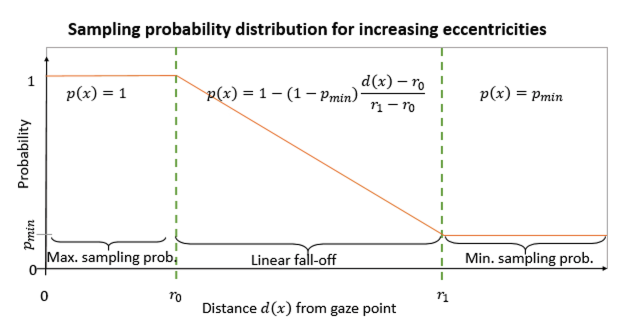}
\caption{Linear acuity model used for sampling pixels in the region between $r_0$ and $r_1$. Image adapted from Weier~\etal~\shortcite{raytracing_hmd}.}
\label{Linear fall off}
\end{figure}

Weier~\etal~\shortcite{raytracing_hmd} consider a linear fall-off of acuity with increasing eccentricity as opposed to the above hyperbolic fall-off. By modeling a linear model into the ray-tracing pipeline, the sampling probability reduces with an increase in distance from the gaze point. 
The visual field is divided into three regions, similar to  Guenter~\etal~\shortcite{guenter2012foveated}. Each region is characterized by three parameters: $r_0, r_1, p_{min}$ as shown in Figure~\ref{Linear fall off}. The inner region (the region which falls within $r_0$ degrees) is the foveal area, which is rendered at full resolution and so is sampled with a probability of one. The pixels in the outermost peripheral region (the region that lies beyond $r_1$ degrees from the gaze point) are sampled with a minimum probability of $p_{min}$. The pixels in the region between the two layers (between $r_0$ and $r_1$) are sampled according to the linear equation:
\begin{equation}
p(x) = 1 - (1-p_{min})\frac{d(x)-r_0}{r_1-r_0}, 
\end{equation}
where $d(x)$ is the distance of the pixel $x$ from the center of the visual field in degrees. $r_0$, $r_1$, and $p_{min}$ are user-defined parameters. 
The linear approximation model is more relaxed and holds good only for small eccentricity values, beyond which the region is rendered at a uniform lower resolution.
However, the linear fall-off maintains motion perception in the periphery better than the model with hyperbolic fall-off due to the increasing sampling rate.

\subsubsection{Logarithmic model}
\label{non-linear models}
In the human visual system, a log-polar mapping of the eye's retinal image approximates the excitation process of the visual cortex~\cite{log_mapping_retina}.
This log-polar mapping ensures that the sensitivity to perceive fine details is high at the center of the visual field and decreases logarithmically with an increase in distance from the fovea. 
\begin{sjnew}
Contrasting the above models that use multi-resolution rates during rendering, a mapping from Cartesian space to log-polar space that matches human visual acuity can use a single uniform resolution rendering, applied directly to the transformed space. 
\end{sjnew}
Log-polar mapping has been proven useful in many areas like computer vision, robotics, computer graphics, and image processing because of its power to provide enough visual detail using limited computational resources~\cite{log_polar}.

\begin{figure}[!h]
\centering
\includegraphics[trim= 0cm 10cm 0 0, clip, width=0.45\textwidth]{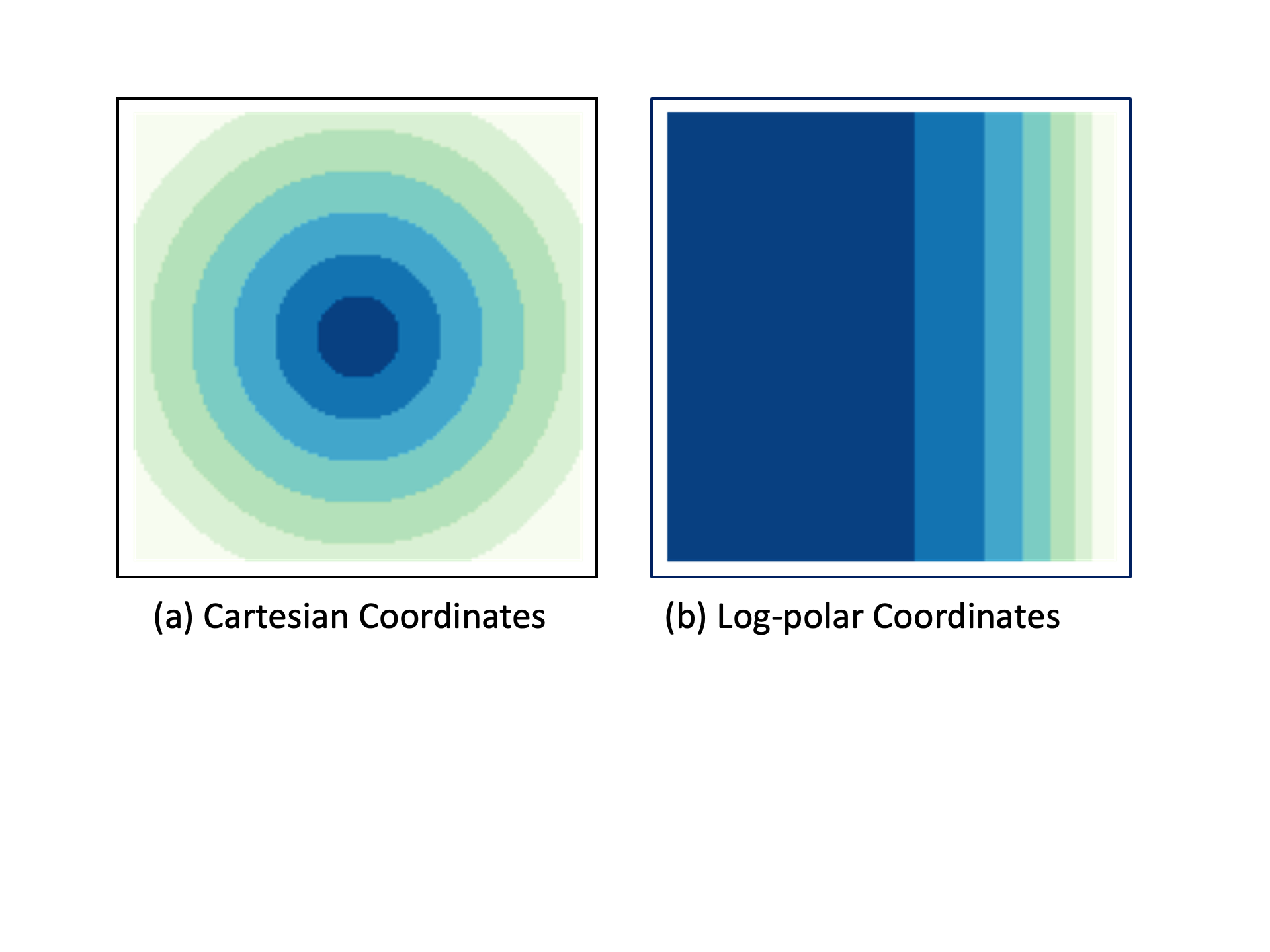}
\caption{Kernel Log-polar mapping to match the human acuity fall-off. The gaze point is at the center of the image. (a) is the image in Cartesian coordinates. (b) is the corresponding image in log-polar coordinates. The region close to the gaze point occupies the major portion in the log-polar space. Image adapted from Meng~\etal~\shortcite{meng2018kernel}.}
\label{Kernel Foveated Rendering}
\end{figure}

Meng~\etal~\shortcite{meng2018kernel} provide a foveated rendering method for meshes using the log-polar mapping of the human visual system. Their system introduces a kernel log-polar mapping technique that offers the flexibility to model acuity fall-off that matches HVS. Figure~\ref{Kernel Foveated Rendering} shows the mapping from the Cartesian space to the log-polar space. The acuity decreases logarithmically and depends on the kernel. The rendering acceleration in this technique is achieved through deferred shading~\cite{deering1988triangle}, a widely-used technique in real-time rendering. The information about the positions, normals, textures, and materials for each surface necessary for shading computations is rendered into the geometry buffer (G-buffer). The contents of the G-buffer are transformed from the Cartesian space to the log-polar space. The direct and indirect lighting at each pixel is computed and rendered to the reduced resolution log-polar buffer. Lighting calculations are performed in the reduced log-polar space. Inverse kernel log-polar mapping is applied to map the shading back to the Cartesian screen space. This method is able to systematically vary the sampling rate and sampling distribution continuously in the log-polar space. 

Koskela~\etal~\shortcite{visual_polar_space} introduce a similar mapping from Cartesian space to visual-polar space. The path tracing with one sample per pixel is performed in the visual-polar space. The polar-coordinate space is modified so that the sampling distribution aligns with the human visual acuity distribution.
To match HVS, the number of samples along the angular and the radial axes can be adjusted.  
They observe that varying the number of samples along the angular axis results in peripheral region artifacts while varying the samples along the radial axis leads to foveal region artifacts. Based on these findings, their optimization technique varies the resolution along the angular axis in the fovea and rescales the radial axis in the peripheral region. 
Denoising is applied to the noisy path-traced visual-polar space image~\cite{koskela2019blockwise, visual_polar_space}.  The reconstructed visual-polar space image is transformed back to the Cartesian coordinates using the inverse mapping. 
They report that the visual-polar mapping reduces distracting artifacts compared to the log-polar mapping.

\subsubsection{Other Approximations : }
Conformal rendering~\cite{bastani2017foveated} models visual acuity as a non-linear function of eccentricity. The technique aims to mimic the smooth transition from the foveal to the peripheral region using a non-linear mapping of the screen distance from the eye-gaze point. The projected vertices of the virtual scene are warped into a non-linear space that matches the retinal acuity and the HMD lens characteristics. The warped image is then rasterized at a reduced resolution and unwarped back into Cartesian space. The complexity of conformal foveated rendering depends on the complexity of the scene. As the number of vertices in a scene increases, the performance reduces. In contrast to the methods that use discrete layers to model the acuity fall-off~\cite{guenter2012foveated, raytracing_hmd}, the methods by Meng~\etal~\shortcite{meng2018kernel} and Bastani~\etal~\shortcite{bastani2017foveated} simulate the acuity fall-off from foveal to the peripheral region using a continuous smooth function with non-linear mapping. This smooth transition helps in reducing visual artifacts.

Reddy~\shortcite{reddy_perceptual} proposes a visual acuity model as a function of the angular velocity of a stimulus projected onto the retina and the eccentricity. The visual acuity varies as an inverse quadratic function of eccentricity~\cite{reddyb}.
Zheng~\etal~\shortcite{perceptual_model_optimized_efficient_fr} develop a foveated rendering method based on the visual acuity model by Reddy~\shortcite{reddy_perceptual} and adjust the tessellation levels accordingly.
Friston~\etal~\shortcite{perceptual_rasterization} use a simple radial power-falloff function $p(x) = f(x^2)$ that maps the distance from the gaze point to a foveated distance. The pixel locations are then scaled based on this foveated distance. The overall effect magnifies the region close to the gaze point, giving more importance to the foveal region. They assume that the foveation function is arbitrary and free to change every frame, but require the function to be invertible to map the foveated image back to an unfoveated one for display.
Fujita and Harada~\etal~\shortcite{raytracing_vrheadset} assume that the acuity drops off as a function of $d(x)^{-2/3}$, where $d(x)$ is the distance from the gaze point and define a sampling distribution function correspondingly.
\begin{new}
Recently, Li~\etal~\shortcite{li2021logrectilinear} presented a  log-rectilinear mapping-based foveated rendering to model exponential decay in resolution with an increase in eccentricity.




\begin{figure}[!t]
\centering

\includegraphics[trim=60 50 80 50,clip, width=0.9\linewidth]{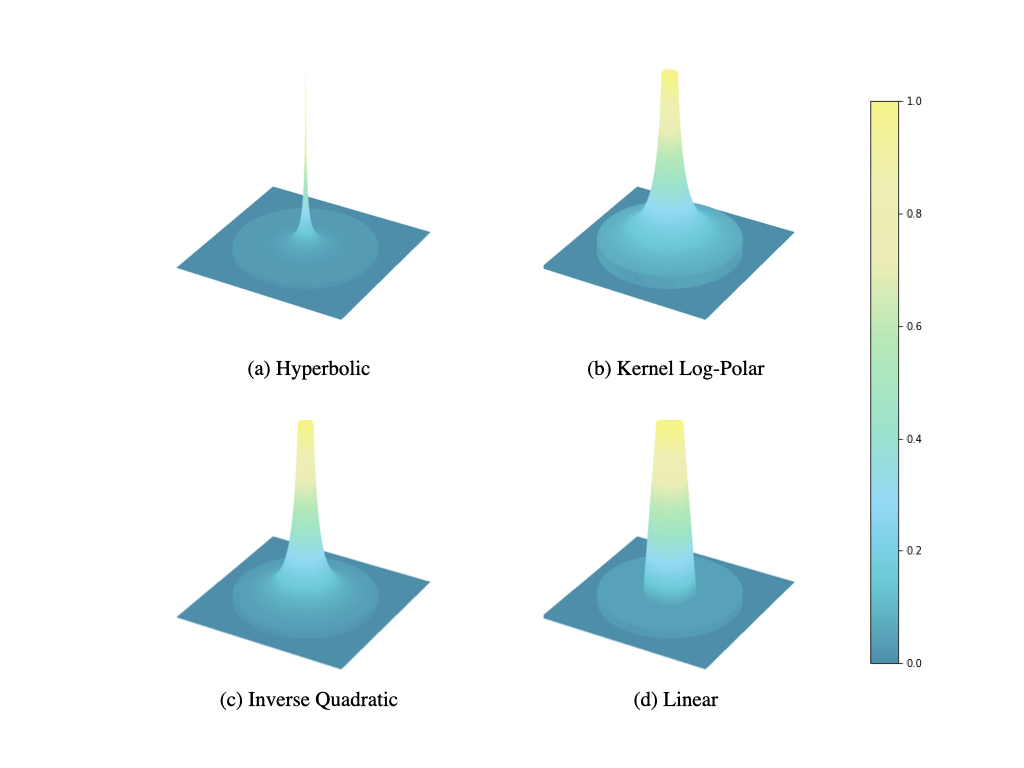}

\caption{Various approximate analytical models are used to vary the spatial resolution across the visual field. We plot the resolution factor as a function of distance from the center of the visual field following (a) hyperbolic fall-off~\cite{guenter2012foveated}, (b) kernel log-polar fall-off~\cite{meng2018kernel}, (c) inverse quadratic fall-off~\cite{reddy_perceptual} and (d) linear fall-off~\cite{raytracing_hmd}.}
\label{fig_distributions}
\end{figure}

The various perceptual models described above mimic the visual sensitivity of the HVS across the visual field. They differ mainly in their motivation, complexity, and distribute samples with subtle differences. 
The most commonly used hyperbolic model mimics the very early operation of retinal projection in the eye, whereas logarithmic model represents the visual cortex part of HVS. 
Computational complexity is a major factor for practical purposes, especially when real-time rendering is required. The least complex, linear model, results in a higher resolution level in the peripheral region compared to the other models. The logarithmic model offers a more tunable fall-off distribution.

Table~\ref{tab:res_models} shows how the visual acuity degrades as a function of eccentricity $e$ for each model.
Futhermore, Figure~\ref{fig_distributions} shows the resolution factor distributions across the visual field.
The parameters are chosen from the corresponding papers. The models mainly differ in the resolution requirements in the near-peripheral region. The hyperbolic function offers a drastic resolution reduction, followed by kernel log-polar, inverse quadratic, and linear fall-offs. 
The kernel log-polar mapping demands higher resolution in the near periphery compared to the hyperbolic fall-off. In the far periphery, all the models converge to similar resolutions. 
One can choose a resolution distribution function from many of these approximations based on the scene complexity and the task to be performed. Moreover, we
have seen that several visual factors impact the perceptual quality of a scene. Most of the existing models consider only a few of these visual factors. There has not been much research thus far in weighing these factors and their inter-relationships to develop an overall perceptual model. Furthermore, the visual sensitivity can change dynamically based on the external stimuli presented to the user, further increasing the complexity of the required model.
Techniques for simplifying the complex models that consider the various inter-relationships between the visual factors without affecting the perceived visual quality or performance are therefore important. 
\end{new}

\begin{table}[]
    \centering
    \begin{tabular}{|c|c|}
         \hline
         Distribution model & Function governing change of resolution\\
         \hline
         Hyperbolic & $1/e$ \\
         Linear & $-e$ \\
         Logarithmic & $1/log(e)$ \\
         Inverse Quadratic & $1/ e^{-2/3}$ \\
         \hline
    \end{tabular}
    \caption{Various models characterizing the degradation of visual sensitivity as a function of eccentricity e}
    \label{tab:res_models}
\end{table}
\subsection{Foveation Space}
\label{Foveation space}

\begin{figure}[h]
\centering
    \includegraphics[trim=250 400 400 400,clip,width=0.6\linewidth]{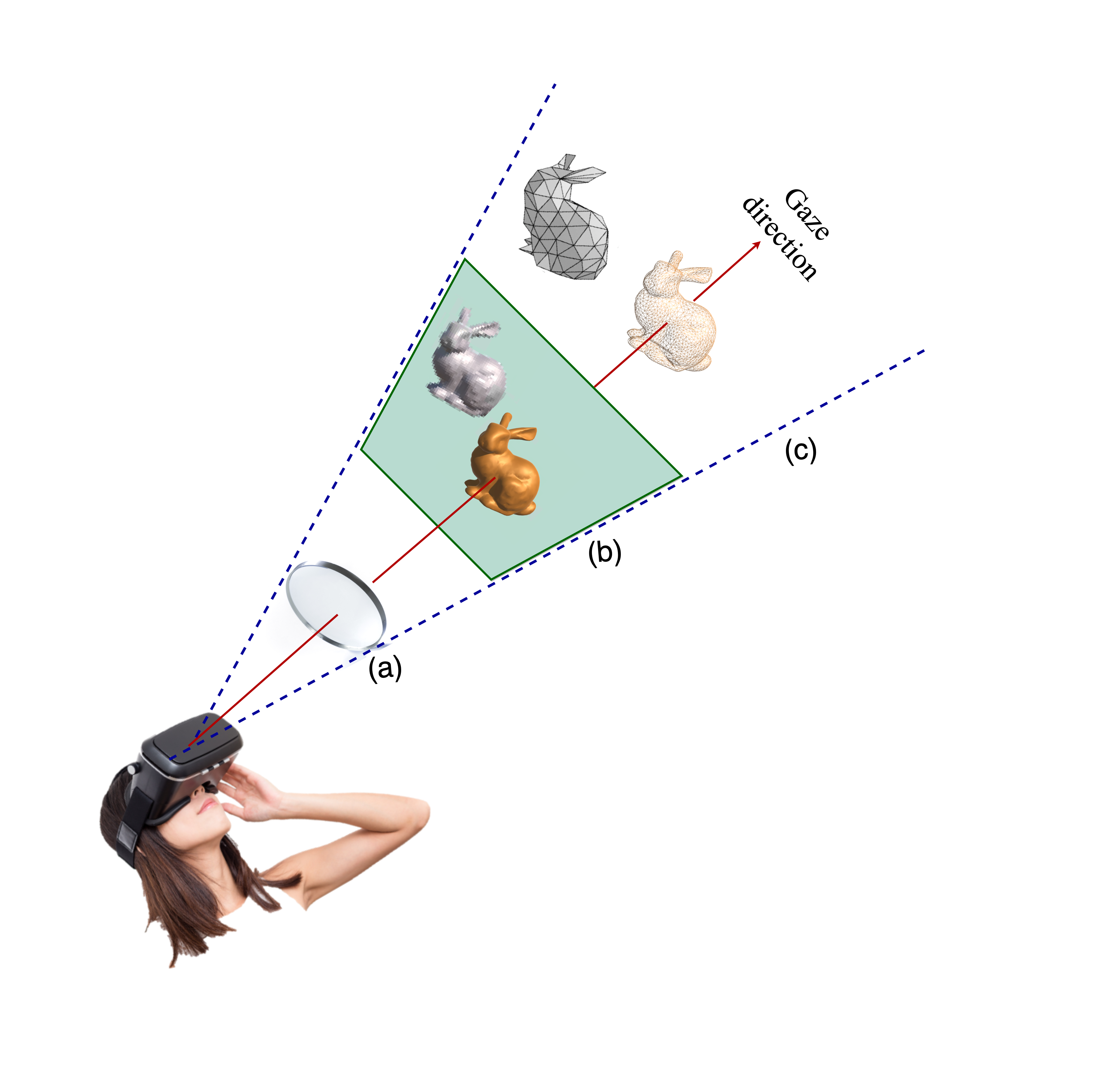}
\caption{Foveation can be applied at any of the three stages - (a) Optics space, (b) Screen space and (c) Object space.}
\label{fig_space}
\end{figure}

\begin{new}
We classify foveated rendering techniques into screen-based, object-based, or optics-based methods based on the stage of the rendering pipeline that incorporates the concept of foveation: the screen space, the object space, or the optical space (Figure ~\ref{fig_space}).
\myremove{Foveated rendering techniques for rasterization can be classified into  screen-based or object-based methods based on the stage of the rendering pipeline that incorporates the concept of foveation.}
\end{new}

\subsubsection{Screen-based Methods}
Most foveated rendering techniques vary the sampling rate in the screen space based on the distance from the point of focus or the gaze point. Screen-based foveated rendering approaches~\cite{guenter2012foveated, patney2016towards, meng2018kernel, vaidyanathan2014coarse, he2014extending} involve manipulating the frame-buffer contents just prior to the display to reduce the overall shading rate. 
Vaidyanathan~\etal~\shortcite{vaidyanathan2014coarse} provide an architecture called coarse pixel shading that enables sparse shading operations for the peripheral region in screen space.
Swafford~\etal~\shortcite{swafford2016user} study the effect of foveation on ambient occlusion which is associated with a high computational cost on modern real-time rendering pipelines.  Specifically, they present the effect of varying per-pixel depth samples in foveal and peripheral regions for Screen-Space Ambient Occlusion, a technique which approximates ambient occlusion. They show that the banding effect arising from a low number of per-pixel samples is imperceptible in the peripheral region due to the reduced acuity, thus improving performance. 
Another foveated renderer that makes changes in the screen space is conformal rendering~\cite{bastani2017foveated} in which the projected vertices are warped to a non-linear space before rasterization. The image is then rasterized at a reduced resolution and finally unwarped into the Cartesian space.

\subsubsection{Object-based Methods}
Object-based foveated rendering methods involve manipulating the model geometry prior to rendering. One of the very early works on foveated rendering by Levoy and Whitaker~\shortcite{levoy} is an object-based approach. 
Following a ray-tracing approach for volume rendering, the number of rays cast per unit area and the number of samples per unit length along each ray are changed based on the pixel's angular distance from the gaze direction.
The system prefilters the 3D volume using a 3D MIPmap and uses fewer samples in the peripheral region. Several early works~\cite{murphy_lod, ohshima, danforth2000, simplifications_leubke} also model objects based on the gaze direction. The desired level of detail for the object is determined based on its distance from the gaze point. Swafford~\etal~\shortcite{swafford2016user} have developed another object-space technique. They vary tessellation levels in the foveal and peripheral regions. A higher tessellation factor is used for the tiles that will fall within the foveal area after screen-space projection. For those that fall in the peripheral region, a lower tessellation factor is used. 
Linear interpolation determines the level of tessellation for the tiles that fall in the region between the foveal and peripheral areas.
Their results show performance gain compared to rendering the entire scene at a uniform tesselation. Along similar lines to Swafford~\etal~\shortcite{swafford2016user}, Zheng~\etal~\shortcite{perceptual_model_optimized_efficient_fr} propose a method to tessellate based on visual sensitivity at a given eccentricity. 
They suggest removing all the imperceivable polygons and then dynamically tessellating the model. The polygons whose edge length in the screen space is less than the user's minimum perceptible length are not further sub-divided. The tessellation level for the perceivable polygons is given by the ratio of the edge length of the polygon in the screen space to the minimum perceptible length.

\subsubsection{Optics-based Methods}
Optics-based foveated displays involve optically manipulating light from one or more displays~\cite{prescription_ar, geometric_phase_polarization, foveated_ar, lee_liquidcrystalphotonics, yoshida_1995, Rolland_yoshida_OHRI}. Such methods employ additional optical components in the device design, and so the overall form factor and the manufacturing cost of the device increases. 
Foveation effect at optical level can be achieved by either using a single display~\cite{geometric_phase_polarization}, or two separate displays~\cite{foveated_ar, lee_liquidcrystalphotonics, yoshida_1995, two_display_tan}. 
Yoo~\etal~\shortcite{geometric_phase_polarization} propose a single-display-based near-eye foveated system based on temporal polarization multiplexing and provides two operating modes, whereas Kim~\etal~\shortcite{foveated_ar} and Lee~\etal~\shortcite{lee_liquidcrystalphotonics} use two displays -- one for the foveal region with a narrow FOV and the other for the peripheral region with a wide FOV.\\
The concept of double displays dates back to the late 1990s. Double-display systems optically combine the light from two display modules, each with a different resolution and field of view.
Yoshida~\etal~\shortcite{yoshida_1995} propose a high-resolution-inset HMD, which optically superimposes a high-resolution image over a low-resolution wide FOV image. The part of the scene around the gaze point is generated at a high resolution. The high-resolution-inset is optically duplicated into a grid of non-overlapping copies to fill the entire display~\cite{yoshida_1995, Rolland_yoshida_OHRI}. An LCD array selects one element of this grid based on the gaze direction and transports it to the eyes through optical fibers~\cite{yoshida_design}. The background image covers the entire FOV and is generated at a lower resolution. The foveal inset is then combined with the background image using opto-electrical components alone. However, these devices were too heavy and expensive when they were proposed.
With recent developments in the optical elements, Kim~\etal~\shortcite{foveated_ar} and Lee~\etal~\shortcite{lee_liquidcrystalphotonics} present prototypes for foveated near-eye devices using two displays with a small form factor. 
Kim~\etal~\shortcite{foveated_ar} use a micro OLED display for the foveal display system and a projector-based Maxwellian-view display for the peripheral display system. 
The two displays are mechanically steered towards the gaze direction based on the built-in eye tracker's information.
Lee~\etal~\shortcite{lee_liquidcrystalphotonics} use a holographic near-eye display for the foveal display and elements based on polarization optics for the peripheral display system. 
The foveal display is steered using a micro-electro-mechanical system (MEMS) mirror and a switchable Pancharatnam–Berry phase (S-PBP) grating module.
The peripheral display is ensured to support enough eye box, thereby avoiding the need to steer the display. The light coming from these two displays are then combined optically before reaching the eye.\\
The concept of foveation is also increasingly used in holographic displays~\cite{holographic_displays_survey, holographic_NED_maimone}, a highly promising direction to true 3D displays with powerful features such as variable focal control, optical aberration correction, and non-conflicting depth cues, to reduce the rendering cost of computer-generated holograms (CGH)~\cite{Foveated_CGH_layer_chang, Foveated_CGH_mesh_Ju, Foveated_CGH_ray_tracing_sakamoto}. 
For example, Chang~\etal~\shortcite{Foveated_CGH_layer_chang} use the concept of foveated rendering to speed-up the computational cost of CGH for 3D volumes based on a layered approach. 
The volume is divided into several parallel layers, and the image at each layer is segmented into a foveal region and a peripheral region. 
The foveal region is considered at a high resolution for the hologram calculation, whereas the peripheral region is down-sampled to a lower resolution. As a result, the reconstructed images from the hologram appear with a higher quality in the foveal region than the peripheral region.

\subsection{Gaze Point Information}
\label{Static vs Dynamic}
\begin{figure}
\centering
\includegraphics[width=0.23\textwidth]{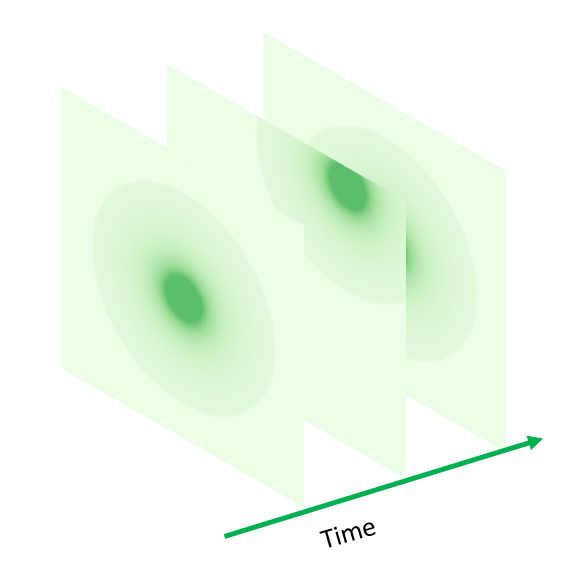}
\includegraphics[width=0.23\textwidth]{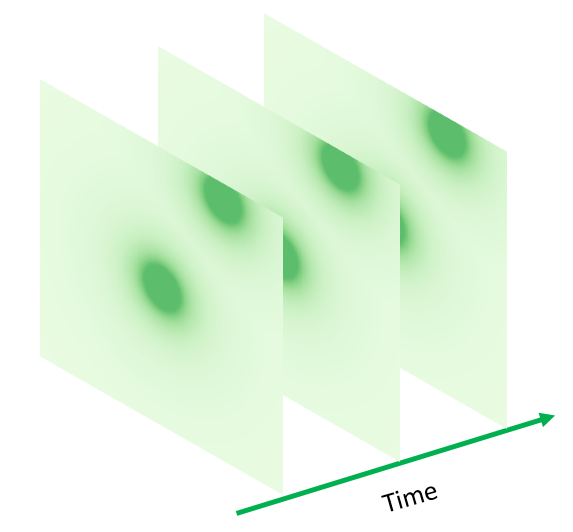}\\
\hspace{0.01\textwidth} \text{Dynamic} \hspace{0.2\textwidth} \text{Static}
\caption{Dynamic and Static Foveated Rendering. Dynamic foveated rendering depends on the information from the eye-tracking hardware. Static foveated rendering does not rely on the eye-tracking hardware.}
\label{fig_static_dynamic}
\end{figure}

\begin{new}
This section discusses the use of eye-tracking knowledge for foveated rendering systems (Figure~\ref{fig_static_dynamic}) and classifies the methods based on the configuration in which they were originally proposed. 
\end{new}

\subsubsection{Static Foveated Rendering}
Static foveated rendering techniques do not rely on the eye-tracking devices and assume that a user focuses on specific positions in the image. 
The likelihood of a user glancing at a particular area estimates the desired acuity level in that region.
Many early works exploited static foveation, in which they either assume that the user looks at the center of the screen~\cite{funkhouser1993adaptive} or use a content-based model of visual attention~\cite{horvitz1997perception, yee2001spatiotemporal}. The direction of the head was considered a good approximation for the gaze direction before the eye-tracking devices were developed. 
\begin{new}
Most of the early work on foveated rendering is based on this approximation and assumes that the gaze point is at the center of the screen. 
\myremove{The early works on foveated rendering are based on this approximation and assume that the gaze point is at the center of the visual field.}
\end{new}

\begin{new}
One example of static foveated rendering is the fixed foveated rendering (FFR)~\cite{oculus_ffr} developed by Oculus, which assumes that most users look towards the center of the display.
\myremove{Fixed foveated rendering (FFR)~\cite{oculus_ffr} also assumes that users look towards the center of the display.}
\end{new}
FFR uses a tile-based approach. The image is sub-divided into tiles of varying acuity levels, based on the distance from the center.
The tile corresponding to the higher resolution region lies at the center while the tiles towards the edges correspond to lower resolution. Each tile is rendered at a uniform spatial resolution. Static foveated rendering requires the users to maintain their focus at predefined fixed zones, mostly the center of the screen. A significant advantage of the static foveated rendering techniques is that they do not depend on the quality of the eye-tracking devices and are compatible with all existing devices.
Moreover, static foveated rendering approaches can be further improved with the recent improvements in the visual attention and saliency models~\cite{saliency_based_visual_attention, mesh_saliency, mesh_saliency_eye_fixations}.
The area around the salient regions is always rendered at a higher resolution. However, static techniques require a larger high-resolution region than the dynamic approaches. So the average rendering cost for a static foveated rendering technique could be higher than the dynamic method.

\subsubsection{Dynamic Foveated Rendering}
Traditional foveated rendering is predicated on eye-tracking technology~\cite{reddy_perceptual}. Eye-tracking devices can pinpoint the user's gaze position in real-time. Research on head-mounted eye trackers dates back to the 1960s~\cite{eye_tracking}. However, the recent advancements in computational power, massively parallel image processing, low cost, and small-sized hardware have made it possible to use real-time eye-tracking with VR and AR.
By actively tracking the user's gaze using eye-tracking tools, a small region around the gaze point is rendered at a higher resolution and the peripheral region at a lower resolution. Most of the foveated rendering approaches~\cite{guenter2012foveated, patney2016towards, swafford2016user, tursun2019luminance, stengel2016adaptive} are dynamic approaches and depend on the performance of the eye-tracking devices to obtain the gaze point. 
In addition to the hardware-based eye-trackers, there has been a growing interest in developing methods to predict future gaze positions using deep-learning methods~\cite{Xu_gaze_2018_CVPR, DGaze}. As the accuracy and efficiency of the eye-tracking solutions improves further, dynamic foveated rendering is likely to greatly enhance the rendering performance and quality.

\subsection{Anti-aliasing}
\label{Anti-Aliasing}

Aliasing is prominent in foveated rendering as the peripheral region is rendered at a lower resolution.
The aliasing artifacts stemming from foveated rendering can be either spatial or temporal.

Spatial aliasing artifacts occur when the level of detail of the virtual world is higher than the rendered resolution~\cite{hoffman201865_displaysystems}. 
They arise at the object level and appear irrespective of any motion.
Aliasing during scene motion generates temporal artifacts. These artifacts are aligned to the output display pixel grid rather than the virtual world coordinates~\cite{hoffman201865_displaysystems}. 
As the user view changes, these artifacts result in flickering and scintillation effects, disrupting the user experience in the virtual world. 

Studies show that the human visual system is sensitive to temporal aliasing artifacts even at higher eccentricities~\cite{hoffman201865_displaysystems, hoffman2018limits}.
McKee and Nakayama~\shortcite{mckee1984detection} show that motion acuity falls dramatically from $0$\textdegree\ to $10$\textdegree\, and then drops more subtly from $10$\textdegree\ to $40$\textdegree~.
The peripheral motion sensitivity of the human visual system poses a critical challenge to all foveated rendering techniques.
Patney~\etal~\shortcite{patney2016towards} show that minimizing temporal aliasing in foveated rendering is necessary for an effective user experience.

\subsubsection{Avoid Artifacts}
\label{Avoid Artifacts}
\begin{figure}[!h]
\centering
\includegraphics[trim=10cm 10cm 10cm 10cm, clip, width=0.45\textwidth]{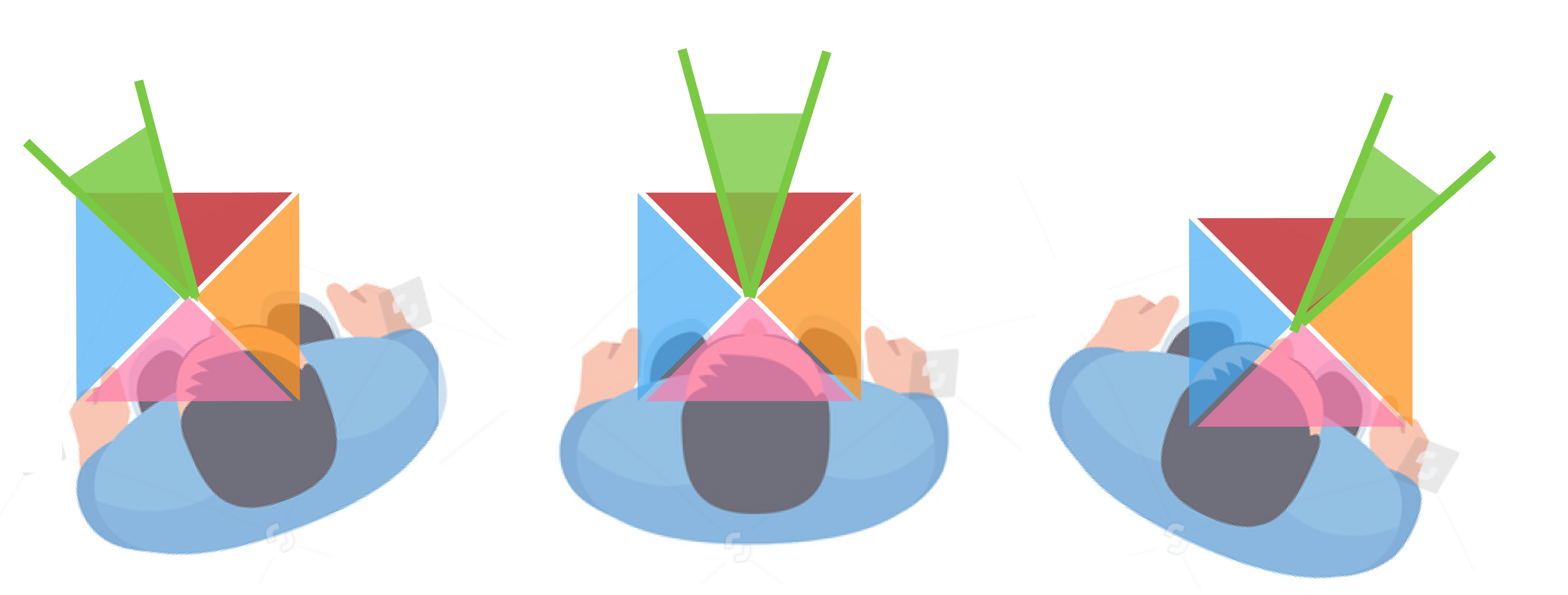}
\caption{Phase-aligned Foveation. The foveal region (green) is rendered in display coordinates, while the peripheral regions (red, orange, pink and blue) are rotationally fixed to the world coordinate system. Image adapted from Turner~\etal~\shortcite{eric2018phase}.}
\label{fig-Phase-aligned Foveated Rendering}
\end{figure}

One approach to ensure temporal and spatial stability is to design foveated rendering techniques that prevent artifacts from occurring.
These techniques generally identify the feature that might cause a detectable artifact and remove that feature or nullify its effect in the rendering pipeline~\cite{bastani2017foveated}. 
The foveal and peripheral regions are constantly updated according to the gaze direction.
In general, the position of the rendered pixels in the high-resolution foveal region and the low-resolution peripheral region aligns with the display-coordinate system. 
The low-resolution rendered pixels are later upsampled to match the native display resolution.
As the user's head rotates, the value of each rendered pixel changes irrespective of the scene content, causing the pixel color to shift and flicker. 
This generates the time-varying aliasing artifacts in the upsampled display pixels, thus disrupting the user experience.
Turner~\etal~\shortcite{eric2018phase} observe that proper angular alignment of the rendered frustums minimizes such frame-to-frame flickering effects with head rotation.
They present a phase-aligned foveated rendering system that aligns the low-resolution region to the world-coordinate system rather than the display-coordinate system as shown in Figure~\ref{fig-Phase-aligned Foveated Rendering}.
The low-resolution region, which is sampled in the world space, is then upsampled and re-projected to align with the display-coordinate system.
The artifacts in the peripheral region now move along with the content of the virtual scene, and so the temporal variation of the artifacts is minimized as the head rotates.
Phase-aligned foveated rendering significantly reduces the perceivability of motion artifacts in the peripheral region, making aggressive foveation possible. 
However, spatial aliasing and a distinct boundary between the foveal and peripheral regions are noticeable in the rendered frame. The discern-ability of the foveal-peripheral boundary can be reduced by ensuring a smooth transition in the resolution levels from foveal to the peripheral region~\cite{bastani2017foveated}. 
{Franke~\etal~\shortcite{time_warped} and Mueller~\etal~\shortcite{temporally_adaptive_shading} use temporal coherence to reduce the number of shading operations. 
Franke~\etal~\shortcite{time_warped} reproject the peripheral region of the previous frame to the current frame, and only the foveal pixels and the less coherent peripheral pixels are rendered. They propose to use the exact world space pixel positions of the fragments rather than the depth values to avoid reprojection artifacts.}

\subsubsection{Mitigate Artifacts}
\label{Mitigate Artifacts}

Most of the foveated rendering approaches reduce the perceptibility of the artifacts in the post-rendering phase.
The choice of interpolation techniques used to up-sample the low-resolution peripheral region also affects the detection of the artifacts in the peripheral region.
For example, the simple nearest-neighbor interpolation magnifies the temporally unstable artifacts~\cite{hoffman2018limits}, though it can offer good contrast retention when compared to bi-linear interpolation.

Guenter~\etal~\shortcite{guenter2012foveated} use a combination of multi-sample anti-aliasing (MSAA), temporal reverse reprojection, and temporal jitter of the spatial sampling grid to reduce the spatial and temporal artifacts. MSAA reduces spatial aliasing along silhouette edges by increasing the effective sampling resolution in those regions. Temporal reverse projection with frame jitter reduces the aliasing throughout the image. 
Temporal anti-aliasing strategies are better at reducing aliasing arising from sampling highly specular materials at a lower sampling rate. 
Reprojection-based temporal anti-aliasing is a common technique that uses the previous frame's information to mitigate the temporal artifacts. The current frame is reprojected onto the previous frame, and the information from the two frames is used to compute the final color. 
However, as details from the previous frame may continue to exist beyond when they can be correctly reprojected, temporal anti-aliasing may produce high-frequency artifacts or ghosting~\cite{patney2016towards}. 
Karis~\shortcite{karis2014high} reduces such ghosting artifacts by conditioning samples from previous frames that are consistent with samples in the current frame. Meng~\etal~\shortcite{meng2018kernel} apply temporal anti-aliasing with Halton sampling in the screen space as a post-processing method to mitigate the artifacts appearing in the peripheral region.

Patney~\etal~\shortcite{patney2016towards} use pre-filters in addition to temporal anti-aliasing methods to mitigate temporal artifacts. 
They introduce variance sampling as a post-process image-enhancement technique.
Around each pixel, a variable-size axis-aligned bounding box is constructed based on local color distribution.
The back-projected and resampled information from the previous frame that lies within the defined bounding box is integrated with the current frame information.
The authors show that by explicitly using the local color information, the ghosting artifacts are reduced.
This method is further extended to account for saccadic eye movements. 
Due to the saccadic movement of the eye, the previous frame's peripheral region can become the foveal region for the current frame.
Such a situation requires information from several frames to converge to the level of detail required for the foveal region. 
Patney~\etal~\shortcite{patney2016towards} accelerate the rate of convergence based on the shading rates in the two consecutive frames.
This reduces the blurring artifacts caused by eye saccades.

Weier~\etal~\shortcite{weier2018foveated} use depth-of-field information to design a post-process anti-aliasing technique. 
The depth-of-field effect that occurs when focusing on objects can be used as a low-pass filter to minimize the high-frequency artifacts in the peripheral region. 
Temporal anti-aliasing is first applied to the foveated rendered image to obtain temporally smooth samples. 
These samples are used to reconstruct the full image using push-pull interpolation.
The aliasing artifacts are then mitigated using a low-pass depth-of-field filter.
The approximate depth of the focal point is estimated using a support vector machine-based gaze depth estimator that takes various depth measurements as input. 
Given the gaze-depth and estimation inaccuracy, two circles of confusion are computed, which determine the depth-of-field.
A multi-layer filter is designed based on the depth-of-field model. 
The filtering of the image occurs in layers, which are eventually blended with different weights to give the final output.

In addition to the above methods, denoising approaches based on deep neural networks have been developed to reduce the detection of artifacts. 
DeepFovea~\cite{deepfovea} takes inspiration from the internal model of the human visual system that infers content from the sparse peripheral information. A manifold representing the distribution of samples from a large collection of natural videos is learned using generative adversarial networks~\cite{GAN}.
For a given sparse foveated input video stream, DeepFovea reconstructs the peripheral region by finding the closest natural video that corresponds to the sparse input on the learned manifold. 
As the reconstructed output is close to the realistic videos, the temporal aliasing artifacts are minimized.

\begin{table*}[!t]
\renewcommand{\arraystretch}{1.5}
\caption{Classification of existing prominent foveated rendering works based on the presented taxonomy}
\label{tab:papers}
\centering

\resizebox{\textwidth}{!}{
    \begin{tabular}{|l||c|c|c||c|c|c|c||c|c|c||c|c||c|c|}
    \hline
         \textbf{Paper} & \multicolumn{3}{c||}{\textbf{Visual Factors}} & \multicolumn{4}{c||}{\textbf{Distribution}} & \multicolumn{3}{c||}{\textbf{Space}} & \multicolumn{2}{c||}{\textbf{Gaze}} & \multicolumn{2}{c|}{\textbf{Aliasing}} \\
        \hline
        & \rotatebox{90}{Acuity} & \rotatebox{90}{Contrast} & \rotatebox{90}{Saliency} & \rotatebox{90}{Hyperbolic} & \rotatebox{90}{Linear} & \rotatebox{90}{Logarithmic} & \rotatebox{90}{Quadratic} & \rotatebox{90}{Object} & \rotatebox{90}{Screen} & \rotatebox{90}{Optics} & \rotatebox{90}{Static} & \rotatebox{90}{Dynamic} & \rotatebox{90}{Avoid} & \rotatebox{90}{Mitigate} \\
        
        \hline
        
        Funkhouser and S{\'e}quin~\shortcite{funkhouser1993adaptive} & \cmark & - & \cmark & \cmark & - & - & - & \cmark & - & - & \cmark & - & - & - \\
        \hline
        
        Leubke and Hallen ~\shortcite{simplifications_leubke} & \cmark & \cmark & \cmark & - & - & - & - & \cmark & - & - & - & \cmark  & - & - \\
        \hline
        
        Murphy and Duchowski~\shortcite{murphy_lod} & \cmark & - & - & - & - & - & - & \cmark & - & - & - & \cmark & - & - \\
        \hline
        
        Reddy~\shortcite{reddy_perceptual} & \cmark & - & - &
        - & - & - & \cmark & 
        \cmark & - & - & 
        \cmark & - & - & - \\
        \hline
        
        Guenter~\etal~\shortcite{guenter2012foveated} & \cmark & - & - &
        \cmark & - & - & - &  
        - & \cmark &  - & 
        - & \cmark & - & \cmark \\ 
        \hline
        
        Fujita and Harada~\shortcite{raytracing_vrheadset} & \cmark & - & - &
        \cmark & - & - & - &  
        - & \cmark &  - & 
        \cmark & - & - & \cmark \\ 
        \hline
        
        Stengel~\etal~\shortcite{stengel2016adaptive} & \cmark & \cmark & \cmark &
        \cmark & - & - & - &  
        - & \cmark &  - & 
        - & \cmark & \cmark & \cmark \\ 
        \hline

        Swafford~\etal~\shortcite{swafford2016user} & \cmark & - & - &
        \cmark & - & - & - &  
        \cmark & \cmark &  - & 
        - & \cmark & - & - \\ 
        \hline
        
        Patney~\etal~\shortcite{patney2016towards} & \cmark & \cmark & - &
        - & \cmark & - & - &  
        - & \cmark &  - & 
        - & \cmark & - & \cmark \\ 
        \hline
        
        Weier~\etal~\shortcite{weier2017perception_driven} & \cmark & - & - &
        - & \cmark & - & - &  
        - & \cmark &  - & 
        - & \cmark & \cmark & \cmark \\ 
        \hline
        
        Turner~\etal~\shortcite{eric2018phase} & \cmark & - & - &
        - & - & - & - &  
        - & \cmark &  - & 
        - & \cmark & \cmark & - \\ 
        \hline
        
        Zheng~\etal~\shortcite{perceptual_model_optimized_efficient_fr} & \cmark & - & - &
        - & - & - & \cmark &  
        \cmark & - &  - & 
        - & \cmark & - & - \\ 
        \hline
        
        Meng~\etal~\shortcite{meng2018kernel} & \cmark & - & - &
        - & - & \cmark & - &  
        - & \cmark &  - & 
        - & \cmark & - & \cmark \\ 
        \hline
        
        Weier~\etal~\shortcite{weier2018foveated} & \cmark & - & - &
        - & \cmark & - & - &  
        - & \cmark &  - & 
        - & \cmark & - & \cmark \\ 
        \hline
        
        Koskela~\etal~\shortcite{visual_polar_space} & \cmark & - & - &
        - & - & \cmark & - &  
        - & \cmark &  - & 
        - & \cmark & \cmark & \cmark \\ 
        \hline
        
        Tursun~\etal~\shortcite{tursun2019luminance} & \cmark & \cmark & - &
        \cmark & - & - & - &  
        - & \cmark &  - & 
        - & \cmark & \cmark & - \\ 
        \hline
        
        Kim~\etal~\shortcite{foveated_ar} & \cmark & - & - &
        - & - & - & - &  
        - & - & \cmark & 
        - & \cmark & - & \cmark \\ 
        \hline
        
        Lee~\etal~\shortcite{lee_liquidcrystalphotonics} & \cmark & - & - &
        - & - & - & - &  
        - & - & \cmark & 
        - & \cmark & - & \cmark \\ 
        \hline
        
        Yoo~\etal~\shortcite{geometric_phase_polarization} & \cmark & - & - &
        - & - & - & - &  
        - & - & \cmark & 
        - & \cmark & - & \cmark \\ 
        \hline

    \end{tabular}}
         
\end{table*}

\section{Evaluation}
\label{Evaluation}
\begin{new}
One can measure the performance of a foveated system across two dimensions --
the quality of the rendered image and the average rendering time. 
However, quality evaluation is a challenging task as a ground-truth reference foveated image does not exist. 
The peripheral region is sensitive to flickering artifacts, spatial and temporal motion, contrast, and salient features~\cite{patney2016towards} and it is not easy to quantify these perceptual metrics. Further research is required to understand the important characteristics of human perception. 
Also, the optical components of the HMD limit the perceived resolution. The spatial resolution of optical systems is commonly expressed in terms of the modulation transfer function (MTF), which gives the normalized frequency response of the system. Beams~\etal~\shortcite{beams_ang_spatial_res_displays} show that the maximum resolution provided by the optical system depends on the eccentricity in the FOV. This dependence on the optical components also has to be considered while evaluating the foveated rendering system.
This section gives an overview of the common qualitative and quantitative measures used to evaluate a foveated system.
\end{new}

\vspace{-0.2cm}
\subsection{User study-based evaluation}
\label{qualitative}
\begin{new}
Empirical user studies are the most common and reliable way to evaluate foveated rendering systems.
\myremove{
Evaluating the quality of foveation is a challenging task as a ground-truth reference foveated image does not exist. 
The experiments by Patney~\etal~\cite{patney2016towards} show that contrast sensitivity and temporal stability have to be preserved for a good-quality foveated rendering.
Further experiments are needed to understand the features that maintain the perceptual quality similar to that of the human visual system.
As described in Section~\ref{Visual Attributes}, the peripheral region is sensitive to flickering artifacts, spatial and temporal motion, contrast, and salient features. 
All these properties of the peripheral region of the human vision need to be considered when designing an evaluation metric to determine the quality of the foveated image.\\
Traditional image quality metrics are designed for uniform-quality images. Structural Similarity Index (SSIM), HDR Visual Difference Predictor (HDR-VDP2) are well-known perceptually informed metrics~\cite{swafford2016user}. 
SSIM~\cite{ssim} uses structural distortion in an image as an estimate of perceived visual distortion. It is based on the assumption that HVS is highly adapted for the extraction of structural information from the visual field. 
SSIM uses mean, variance, and covariance of the original high-resolution image and the foveated image to measure their structural differences. HDR-VDP2~\cite{hdr-vdp2} considers the contrast sensitivity measurements. 
These metrics were modeled with the assumption of a uniform acuity across the visual field. 
Swafford~\etal~\cite{swafford2016user} extend the HDR-VDP2 metric to consider the falloff in visual acuity with increasing eccentricity. 
However, all the metrics mentioned above require a full-resolution reference image to determine the quality of the foveated image.
As a standard evaluation metric that considers all the spatial and temporal aspects of human vision is not available, current evaluation of foveated rendering relies on empirical user studies. 
The optimal parameters for the foveated rendering system are estimated based on the responses obtained from user-study participants. }
\end{new}
These user studies can be broadly classified into three categories~\cite{frperceivablechih}:
\begin{itemize}
  \item single-stimulus absolute category rating or pairwise comparisons
  \item double-stimulus quality comparison
  \item slider methods or adjustment methods
\end{itemize}
\textbf{Single-Stimulus Absolute Category Rating}: The participant is presented with a single stimulus, a foveally-rendered image, on the screen for a certain time. The participant is then asked to make a judgment on the quality of the image, ranging from acceptable to unacceptable.

\textbf{Double-Stimulus Quality Comparison}: This type of experiment requires the availability of a full-resolution image. Two renderings of the scene, a full-resolution (unfoveated) image, and a foveated image are shown to the participant in a random order. The participant is then asked to compare the two images and determine the perceptually superior image.

\textbf{Slider Methods}: These methods are more helpful in finding the optimal parameters for the foveated-rendering system. There are two types of slider experiments - descending and ascending. In descending methods, the participant is presented with a full-resolution high-quality image and a slider. The slider is used to control the parameters of foveation. By moving the slider position gradually, the participant is asked to determine the point at which the quality of the image degrades. The ascending method is similar to the descending method, but the participant is initially provided with a reference full-resolution image for a short period. The quality of the image gradually increases from a fully foveated image as the slider moves, and the user is asked to determine the point at which the quality of the image is perceptually similar to the reference image.

The user studies also provide personalized calibration of optimal foveation parameters since the level of acceptance of foveated imagery may vary from person to person. 
However, user studies are generally very time-consuming and expensive. 
Therefore, perception and foveation-based evaluation metrics
\myremove{need to be developed}are required to help us design and compare various methods. The user-study experiments can help us evaluate methods that have a high probability of success based on the initial evaluation metrics.

\subsection{Computational metric-based evaluation}
\label{quantitative}
\begin{move}

Traditional image quality metrics are designed for uniform quality images. Structural Similarity Index (SSIM) and HDR Visual Difference Predictor (HDR-VDP2) are well-known perceptually-informed metrics~\cite{swafford2016user}. 
SSIM~\cite{ssim} uses structural distortion in an image as an estimate of perceived visual distortion. 
It is based on the assumption that HVS is highly adapted for the extraction of structural information from the visual field.
SSIM uses mean, variance, and covariance of the original high-resolution image and the foveated image to measure their structural differences. 
HDR-VDP2~\cite{hdr-vdp2} considers the contrast sensitivity measurements. 
These metrics were modeled with the assumption of a uniform sensitivity across the visual field. 
\end{move}
\begin{new}
Several recent papers~\cite{swafford2016user, metric_foveated_quantization_resolution, metric_foveated_receptive_size, metric_foveated_wavelet, metric_foveated_wavelet_attention, metric_foveated_mse_video, lee_fwsnr_fpsnr} extend the uniform image quality metrics to consider foveation.
Rimac~\etal~\shortcite{rimac2011foveation} combine SSIM with a foveation-based sensitivity function to account for the non-uniform perceived structural differences across the visual field.
Swafford~\etal~\shortcite{swafford2016user} extend the HDR-VDP2 metric by including the contrast-sensitivity degradation factor.
\myremove{Swafford~\etal~\cite{swafford2016user} extend the HDR-VDP2 metric to consider the fall-off in visual acuity with increasing eccentricity.}
Tsai~\etal~\shortcite{metric_foveated_receptive_size} develop a window-based metric and assign weights based on the distance from the predicted salient regions. Guo~\etal~\shortcite{metric_foveated_quantization_resolution} vary the spatial resolution based on eccentricity to calculate the quality of the image. 
{Recently, Mantiuk~\etal~\shortcite{fovvideoVDP} developed a novel metric named FovVideoVDP that incorporates several factors such as spatial resolution variance with eccentricity, spatial and temporal contrast sensitivity, and scene content.}
Most of these metrics have been developed for foveated video quality assessment and rely on a full-resolution reference image or video. 
However, complete perceptually informed and computationally simple metrics to evaluate the perceived quality of wide FOV images and videos and level of immersion are relatively unexplored.

\end{new}
\myremove{
However, all the metrics mentioned above require a full-resolution reference image to determine the quality of the foveated image.
}

In terms of rendering time, 
we provide the results reported for each of the discussed foveated rendering methods. This, however, should not be interpreted to be a relative comparison among the different methods unless explicitly noted otherwise; the overall rendering time varies with the complexity of the scene, shading models, and the hardware used.
Guenter~\etal~\shortcite{guenter2012foveated} report an overall speedup of $5\times$ -- $6\times$ in rendering time and $10\times$ -- $15\times$ in the number of pixels rendered.
Multi-rate shading~\cite{he2014extending} provides an average speedup of about $3\times$ -- $5\times$ and the multi-resolution approach~\cite{nvidia-multiresolution} provides a speedup of $1.3\times$ -- $2\times$.
Zheng~\etal~\shortcite{perceptual_model_optimized_efficient_fr} reduce the rendering time by a factor of $1.2\times$ compared to Guenter~\etal~\shortcite{guenter2012foveated}. Patney~\etal~\shortcite{patney2016towards} report $2\times$ more efficiency in reducing shading cost compared to Guenter~\etal~\shortcite{guenter2012foveated}.
Stengel~\etal~\shortcite{stengel2016adaptive} report a speedup of $1.34\times$ in overall rendering time while reducing the shading time by $1.7\times$. 
Swafford~\etal~\shortcite{swafford2016user} reduce the rendering time in half by carefully reducing the resolution in the periphery. 
The kernel-foveated rendering method by Meng~\etal~\shortcite{meng2018kernel} achieves $2.8\times$ -- $3.2\times$ speedup by using log-polar transformations. 
By estimating the maximum resolution required for each pixel, Tursun~\etal~\shortcite{tursun2019luminance} report $3.08\times$ speedup in overall rendering time. 
The ray-tracing-based foveated rendering method by Weier~\etal~\shortcite{raytracing_hmd} provides an average speedup of $2.55\times$, reducing the number of sampled pixels by 79\% on benchmark scenes.
By using depth-of-field filtering as a post-processing step, Weier~\etal~\shortcite{weier2018foveated} reduce the number of sampled pixels by 69\% without any visible artifacts.

\section{Conclusion}
\label{Summary}
In conclusion, we have described the limitations of the human visual system and how foveated rendering leverages such limitations to reduce the computational resources required for real-time rendering. 
We compare and discuss various foveated rendering works by providing a taxonomy based on the key factors in the rendering pipeline. 
Table~\ref{tab:papers} provides a summary of our taxonomy. 
Our first classification differentiates methods based on the visual factors considered while developing the rendering system: content-independent (acuity) and content-dependent (image contrast, saliency).
As can be seen from table~\ref{tab:papers}, all the papers consider acuity degradation with eccentricity in their rendering systems. In addition, a few works~\cite{tursun2019luminance, patney_perceptually, stengel2016adaptive} leverage the content-dependent information to further improve the foveation effect.
The second category differentiates the methods based on the underlying analytical models for peripheral degradation. While all the methods reduce the sampling rate with an increase in distance from the focus point, the reduction rate is different across methods. The best-suited model is selected based on the target task and scene complexity. 
We next differentiate methods according to the space where the foveation effect is employed: object-space, screen-space, and optics-space. With the evolution of the graphics pipeline and dedicated hardware, there is an increasing trend to incorporate foveation in the screen or the image space. More recently, several works have modified the display at the lens level to incorporate foveation. These optics-based foveated methods mitigate the hardware limitation in maintaining high pixel densities to match the angular resolution of the HVS. 
We also separate the methods based on the amount of gaze information used in their initial proposed solution. With improvement in eye-tracking technology, more and more works tend to develop and study foveated systems using gaze point information.
As the eye-tracking technology improves in both accuracy and latency, dynamic foveated rendering seems to be the way to achieve real-time rendering with high perceptual quality.
We also discuss anti-aliasing approaches that complement foveated rendering. These include deep-learning-based approaches that produce realistic, anti-aliased, foveated output images from sparse inputs. 
It is to be noted that the provided categorization is not disjoint, and a solution can combine several options in each category, for example, including several visual factors or combining screen-space and optics-space foveation methods.

We also provided a brief overview of the human visual system.
However, the way the human brain processes information from the images formed on the retina is not yet fully understood. As the neuroscience community advances our understanding of visual perception and cognition, their studies can guide our efforts towards minimal and sufficient rendering without compromising human perception and cognition.

The ultimate goal is to build head-mounted displays that can simultaneously provide wide FOV and high resolution that match human perceptual capabilities. 
Visual acuity is also limited by diffraction, aberrations of the eye lens, in addition to the density of the photoreceptor cells~\cite{eye_visual_smith}. In addition, even factors such as refractive error, illumination, contrast, and location of the retina being stimulated affect the perceived image quality.
Studies also show that the photoreceptor density decreases with age~\cite{age_photoreceptors}.
So, these factors are to be considered while studying the effectiveness of a foveated rendering system.
Interestingly, the foveation characteristics differ even between the two eyes. Meng~\etal~\shortcite{eye_dominance_meng} leverages the concept of eye dominance, which states that the human visual system prefers visual stimuli from one eye more than the other. This suggests that the non-dominant eye can permit greater foveation than the dominant eye without any perceptual difference. 
Similarly, given that only rods are active for night vision can be used to render scenes under dim lighting differently than those with bright lighting and at a much lower computational cost. 
Another possible direction of research is to leverage the color discrimination ability of the human eye across the visual field, which decreases with an increase in eccentricity. 
Further, much of the current effort in building foveated rendering system is for a personalized single user system; foveated rendering systems for shared multi-user displays is another exciting area for future research.

A few recent works~\cite{Tan_foveated_display,foveated_ar,lee_ned,foveated_retinal_opt} use the concept of foveation to develop near-eye foveated displays. Kim~\etal~\shortcite{foveated_ar} use a deep-learning-based gaze-tracking system~\cite{kim_gaze_tracking} for pupil center estimation to move the small foveal display appropriately. 
Such systems can also benefit from estimating the future gaze direction based on the current frame and gaze information.
Improving the accuracy of the saliency models based on long-term memory and the information from the current frame can guide the renderer to render appropriate sections at a high resolution. 
As deep-learning models are increasingly able to provide better results in several tasks, they can also be employed to improve the gaze and saliency estimation tasks.

Although there has been significant progress in foveated rendering, it is still challenging to realize the full potential of foveated rendering, especially in complex scenes with photo-realistic lighting.
Further, there is an urgent need to develop a uniform evaluation metric that uses perceptual criteria to quantify the quality of the foveated rendering and allows us to compare various foveation methods.
We believe that the combination of developments in both software and hardware will make foveated rendering more dominant in every commodity headset.


\begin{acks}
This work has been supported in part by the NSF Grants 15-64212, 18-23321, and the State of Maryland's MPower initiative.
The authors would like to thank Alexandar Rowden, Anshul Shah, Sweta Agrawal and Yogesh Balaji for their helpful suggestions.
\end{acks}

\bibliographystyle{ACM-Reference-Format}
\bibliography{6_References}
\end{document}